\documentclass[twocolumn,floatfix]{revtex4}
\usepackage{CJK}
\pdfoutput=1
\usepackage{aurical}
\usepackage[T1]{fontenc}
\RequirePackage{boondox-cal}
\RequirePackage{graphicx}
\RequirePackage{float}
\RequirePackage{adjustbox}
\RequirePackage{graphicx}
\RequirePackage{epstopdf}
\RequirePackage[version=3] {mhchem}
\RequirePackage{amssymb}
\RequirePackage{braket}
\RequirePackage{latexsym}
\RequirePackage{float}
\RequirePackage{subfigure}
\RequirePackage{young}
\RequirePackage[dvipsnames]{xcolor}
\RequirePackage[toc,page]{appendix}
\RequirePackage{bbm}
\RequirePackage{amsmath}
\usepackage{tikz}
\usetikzlibrary{positioning,shapes,fit,arrows,arrows,decorations.text}
\usepackage[utf8x] {inputenc}

\begin{document}

\title{On the microscopic origin of the Interacting Boson Model in the SU(3) limit
}

\author{Andriana Martinou}

\affiliation{ Institute of Nuclear and Particle Physics, National Centre of Scientific Research ``Demokritos'', GR-15310 Aghia Paraskevi, Attiki, Greece.}

\begin{abstract}

A microscopic interpretation of the $\mathcal{s}$ and $\mathcal{d}$ bosons of the Interacting Boson Model is being suggested: the $\mathcal{s}$, $\mathcal{d}$ bosons can be interpreted as symmetric pairs of harmonic oscillator quanta of the valence nuclear shell. Within this interpretation, the $SU(3)$ limit of the Interacting Boson Model results to the Elliott $SU(3)$ symmetry.
\end{abstract}
\maketitle

\section{Introduction}

This work is about the connection of two towering algebraic nuclear models: the Shell Model $SU(3)$ symmetry of Elliott \cite{Elliott1,Elliott2,Elliott3,Elliott4} and the Interacting Boson Model (IBM) of Arima and Iachello \cite{Arima1975,Arima1976,Arima1978b,Scholten1978,Arima1979}. A very insightful historical review about these two models has been written by P. Van Isacker in Ref. \cite{Isacker2011}. 

Elliott was the first to apply symmetries in nuclear physics in 1958, when he introduced the Shell Model $SU(3)$ symmetry (or nowadays called the Elliott $SU(3)$ symmetry) \cite{Elliott1}. This happened 3 years earlier than the first use of symmetries in high energies physics in 1961 \cite{Neeman1961}. With his work Elliott explained how the nucleons in a valence shell, which consists by orbitals with common number of harmonic oscillator quanta, generate the rotational spectrum. Thus he bridged the microscopic picture given by the nuclear Shell Model of Mayer, Haxel, Jensen and Suess \cite{Mayer1,Haxel1949} with the collective and especially with the rotational nuclear properties. Elliott along with Harvey and Wilsdon had applied the Shell Model $SU(3)$ symmetry in the s, d nuclear shell among the harmonic oscillator magic numbers 8-20. This work begun in 1958 with Ref. \cite{Elliott1} and lasted till 1968 with the publication of Ref. \cite{Elliott4}. 

Afterwards in 1975 the idea that the nuclear spectrum can be produced using spherical tensors of degree 0 and 2 (the $s,d$ bosons), was proposed \cite{Arima1975}. This gave rise to the IBM \cite{Arima1976,Arima1978b,Scholten1978,Arima1979}, which supposes that the valence nuclear shell possesses the $U(6)$ symmetry. The $U(6)$ symmetry accommodates three limiting symmetries: the $SU(3)$ symmetry for rotational nuclei, the $U(5)$ symmetry for vibrational nuclei and the $O(6)$ for the $\gamma$- unstable. The connection of the Collective Model of Bohr, Mottelson \cite{Bohr1952} and Rainwater \cite{Rainwater1950} with the IBM has been also studied in Refs. \cite{Dieperink1980,Ginocchio1980a,Ginocchio1980,Bohr1980,Elliott1986}. The Interacting Boson Model, due to its three limiting symmetries, is appropriate for the description of spherical, deformed and $\gamma$-unstable nuclei. 

A question, which raised almost simultaneously with the introduction \cite{Arima1975} of the Interacting Boson Model, was related to the use of bosons employed by it for the description of a fermionic system (the nucleus). Relevant reviews can be found in \cite{Arima1980,Arima1983,Talmi1983,Elliott1985,Iachello1987}.

The approximate description of fermionic systems in terms of bosons, called a boson mapping \cite{Klein1991}, has a long history. Before the introduction of the IBM, two main kinds of boson mappings have been used:\\
 a) The Beliaev-Zelevinsky-Marshalek (BZM) boson expansions \cite{Beliaev1962,Marshalek1971,Marshalek1974,Marshalek1974a,Marshalek1980,Marshalek1981}, in which it is required that the boson images of the fermionic operators preserve the commutation relations of the various operators.\\
b) The Marumori boson expansions \cite{Marumori1964,Marumorib}, in which it is required that the boson images of the fermionic operators preserve the value of the various matrix elements. However, in both methods a BCS (Bardeen-Cooper-Schrieffer) transformation from particles to quasiparticles is performed first, and then the boson mapping is achieved in the quasiparticle basis. The BCS quasiparticle transformation involved, has as a consequence the non-conservation of the number of particles, which makes these approaches inappropriate for the study of nuclei.
 
Various number conserving boson mappings have emerged after the introduction of the IBM. The first to be introduced was the Otsuka-Arima-Iachello (OAI) mapping \cite{Otsuka1978,Otsuka1978b}, in which the fermion space is first truncated by keeping only fermion pairs with total angular momentum $J=0$ and 2, which are subsequently mapped onto $s$ and $d$ bosons, requiring the conservation of the values of the matrix elements, as in the Marumori method \cite{Marumori1964,Marumorib}. Seniority (the number of pairs not coupled to $J=0$) \cite{Rakavy1957,Bes1959,Talmi1962,Talmi1971,Talmi1973,Talmi} plays a crucial role in the OAI mapping, since fermion pairs of a given seniority $v$ are mapped onto states of $d$ bosons with number of $d$ bosons $n_d=v$. Seniority is known to be a good quantum number in the $U(5)$ and $O(6)$ limiting symmetries of the IBM, thus the OAI boson mapping offers a microscopic justification of the IBM approach in vibrational and $\gamma$-unstable nuclei, but not in the $SU(3)$ limit of the IBM, which is suitable for the deformed nuclei \cite{Bonatsos2021a} . 

A boson mapping of a different kind has been introduced by Bonatsos, Klein, and Li (BKL method) \cite{Bonatsos1984,Bonatsos1986,Bonatsos1987,Menezes1989,Menezes1990}. The BKL mapping is radically different from the OAI mapping, since in the BKL case the full fermion space is mapped onto the boson space, thus truncation has to be carried out in the boson space. Furthermore, conservation of the commutation relations is required, as in the BZM method. However, despite these differences, the BZM mapping also uses the seniority basis, thus being inappropriate for the description of deformed nuclei. The basic problem arising in this kind of number-conserving boson mappings is related to the fact that fermion pairs are mapped onto expansions involving many bosons. For a good approximation to be possible, a small parameter has to exist, allowing one to drop terms in the boson expansion beyond a certain order (beyond $s$ and $d$ bosons, for example). However, no such parameter has been found, except in cases of slightly occupied shells, in which only a few particles appear within a large shell, which can accommodate many particles. In such cases, the number of particles over the size of the shell provides a small parameter, but this happens only in vibrational and $\gamma$-unstable nuclei, while in deformed nuclei no small parameter has been found up to date. The only case in which an exact boson mapping of a fermionic system onto $s$ and $d$ bosons is achieved is the Ginocchio model \cite{Ginocchio1979,Ginocchio1980b,Arima1981}, in which a mapping of fermions living in a single $j=3/2$ shell in terms of $s$ and $d$ bosons is achieved. However, the existence of this mapping is not due to the existence of any small parameter, but to the accidental vanishing of a 6-j symbol \cite{Edmonds}.

The serious problems caused by the truncation of the space down to only $J=0$ and $J=2$ pairs have been pointed out by Bohr and Mottelson \cite{Bohr1980,Bohr1982,Bohr1982a}.

Efforts of constructing boson mappings of Shell Model algebras possessing $SU(3)$ subalgebras, which could be appropriate for deformed nuclei, have been made \cite{Bonatsos1985,Bonatsos1986,Bonatsos1986a,Menezes1989a}, but they led to symplectic algebras having $SU(3)$ subalgebras and not to unitary algebras, as the $U(6)$ algebra of the IBM, thus they are irrelevant to the question of providing a microscopic justification of the IBM in its $SU(3)$ limit.

The group theoretical structure of the Collective Model of Bohr and Mottelson \cite{Bohr1952,BohrII}, which is based on the geometrical description of the quadrupole deformation in terms of the two variables $\beta$ (describing the  departure of the nuclear shape from sphericity) and $\gamma$ (corresponding to the departure of the nuclear shape from axiality), has been understood \cite{Afanasev1972} already in 1972, before the introduction of the IBM in 1975 \cite{Arima1975}, as corresponding to a $U(5)$ overall symmetry having an $O(5)$ subalgebra. This result has been corroborated by Chac\'on and Moshinsky \cite{Chacon1976,Chacon1977}, who in addition studied in detail the group theoretical structure of all three limiting symmetries of the IBM \cite{Castanos1979}.  

Following a complementary path, Janssen, Jolos, and D\"onau \cite{Janssen1974} proved in 1974 that the nuclear quadrupole degree of freedom can be described in terms of an SU(6) algebra. However, no $SU(3)$ limiting symmetry was constructed in this case, since no $s$ boson was used in this approach.

In conclusion, while the microscopic justification of the $O(5)$ subalgebra of the IBM, which underlies both the $U(5)$ and $O(6)$ dynamical symmetries of the IBM, has been understood since a long time, as due to the existence of the seniority as a good quantum number in these cases, there has been no justification of equal clarity found for the $SU(3)$ subalgebra. The above observations suggest that the microscopic justification of the IBM in the $SU(3)$ limit requires an approach radically different from the ones employed so far. 

So how is the IBM linked with the Shell Model in the case of deformed nuclei, which arise when the number of particles is comparable with the size of the valence shell? This is the question I shall approach within this article. 

The idea is quite simple: one fills the valence Shell Model space with nucleons and supposes that the Shell Model $SU(3)$ symmetry can be applied there. Therefore the well known antisymmetric wave function of the many fermion problem splits into a spinor and a spatial part. The spatial wave function can be described by the irreducible representations (irreps) $(\lambda,\mu)$ of the Shell Model $SU(3)$ symmetry \cite{Elliott1}.

The spatial part in the Shell Model $SU(3)$ symmetry is a many quanta wave function, not a many particle wave function \cite{proxy5}. The novel idea, which is proposed in this work, is that the {\it symmetric pairs of quanta, which lie in the Shell Model $SU(3)$ wave functions, are the $s$, $d$ bosons of the IBM}. Since the quanta are bosons, the pairs of quanta are bosons too; no approximations or certain circumstances are needed. It will be proven that there are six types of such cartesian,  symmetrized pairs of quanta, which transform into the $s,d$ bosons. Furthermore, the quanta are spherical tensors of degree 1 and so there can only exist two kinds of symmetric pairs of them: spherical tensors of degree 0 and 2. These tensors create the $s$ and $d$ bosons.

\section{The Nuclear Shell Model}\label{Shell}

The Nuclear Shell Model \cite{Mayer1,Haxel1949} is the state-of-the-art theoretical model, which describes the microscopic structure of atomic nuclei. The first assumption of the model is that the protons and neutrons move inside a mean field potential, which may be represented by the three dimensional isotropic harmonic oscillator (3D-HO). Harvey in section 4.2 of Ref. \cite{Harvey} explains in simple words that {\it any} effective nucleon-nucleon interaction can be expanded into terms, out of which the leading term is the harmonic oscillator potential. The second assumption of the Nuclear Shell Model is the existence of a spin-orbit interaction \cite{Mayer1,Haxel1949}, which leads to the prediction of the so called nuclear magic proton or neutron numbers 2, 8, 20, 28, 50, 82, 126, above which large single particle energy gaps appear. This prediction was the major success of the Shell Model. It is accepted that the spin-orbit interaction is a relativistic phenomenon \cite{Duerr1956,Krutov1973,Thies1985,Reinhard1989,Ring1996}, since the kinetic energy of the nucleons in the nucleus approaches the relativistic regime.

The Hamiltonian of a single particle with mass $m$, momentum $p_x,p_y,p_z$, position $x,y,z$ in a 3D-HO potential with frequency $\omega$, in the cartesian coordinate system reads:
\begin{equation}\label{h0}
h_{0}={1\over 2m}(p_x^2+p_y^2+p_z^2)+{1\over 2}{m\omega ^2}(x^2+y^2+z^2)
\end{equation}
The eigenstates of the above Hamiltonian can be expressed either in the cartesian coordinate system $(x,y,z)$ as $\ket{n_z,n_x,n_y}$, or in the spherical coordinate system $(r,\theta,\phi)$ as $\bf\ket{n,l,m_l}$ \cite{proxy4}. The labels $n_z,n_x,n_y$ represent the harmonic oscillator quanta in each cartesian axis obtaining values $0,1,2,...$, while the $n=0,1,2,...$ represents the radial quantum number and the $l,m_l$ stand for the orbital angular momentum and its projection respectively. Notice, that the bold figure kets $\bf \ket{n,l,m_l}$ will be used to distinguish the spherical eigenstates from the cartesian ones $\ket{n_z,n_x,n_y}$ in this article. The total number of the harmonic oscillator quanta for each eigenstate is \cite{Cohen}:
\begin{equation}
\mathcal{N}=n_z+n_x+n_y=2n+l.
\end{equation}

A unitary transformation among the $\ket{n_z,n_x,n_y}$ and the $\bf \ket{n,l,m_l}$ eigenstates has been presented in Ref. \cite{proxy4}. Specifically one may use Eq. (5) of Ref. \cite{proxy4} to transform the eigenstates of the 3D-HO Hamiltonian from the cartesian to the spherical basis and vice versa. For instance for the $p$ shell with $\mathcal{N}=1$ number of quanta the following transformations can be deduced from the conjugate of Eq. (5) of Ref. \cite{proxy4}:
\begin{gather}
{\bf \ket{n,l,m_l}}\rightarrow \ket{n_z,n_x,n_y}:\nonumber\\
{\bf \ket{0,1,-1}}={\ket{0,1,0}-\mathcal{i}\ket{0,0,1}\over \sqrt{2}},\label{m-}\\
{\bf \ket{0,1,0}}=\ket{1,0,0},\label{m0}\\
{\bf \ket{0,1,1}}=-{\ket{0,1,0}+\mathcal{i}\ket{0,0,1}\over \sqrt{2}}\label{m+},
\end{gather}
where $\mathcal{i}$ stands for the imaginary unit.

The operators, which annihilate or create a harmonic oscillator quantum in each cartesian direction, are the \cite{Lipkin}:
\begin{gather}
a_k=\sqrt{m\omega\over 2\hbar}k+{\mathcal{i}\over \sqrt{2m\omega\hbar}}p_k,
a^\dagger_k=\sqrt{m\omega \over 2\hbar}k-{\mathcal{i}\over \sqrt{2m\omega\hbar}}p_k,\label{a}
\end{gather}
with $k=x,y,z$.
The operators of Eq. (\ref{a}) satisfy the boson commutation relations \cite{Lipkin}:
\begin{gather}
[a_k,a^\dagger_{k'}]=\delta_{kk'},\qquad
[a^\dagger_k,a^\dagger_{k'}]=[a_k,a_{k'}]=0\label{b1}
\end{gather}
with $k,k'=x,y,z$.
 The action of the annihilation and creation operators of Eqs. (\ref{a}) on the cartesian eigenstates of the 1D-HO is \cite{Cohen}:
\begin{gather}
a_k^\dagger\ket{n_k}=\sqrt{n_k+1}\ket{n_k+1},
a_k\ket{n_k}=\sqrt{n_k}\ket{n_k-1}\label{a1}
\end{gather}
for $n_k=0, 1, 2, ...$ and $k=x,y,z$.

Inspired from the spherical harmonics $Y_{\mathcal{m}}^{\mathcal{l}=1}$ (appendix A.1 of \cite{Lipas}):
\begin{gather}
Y^1_{-1}\propto{x-\mathcal{i}y\over \sqrt{2}},\qquad
Y^1_0\propto z,\qquad
Y^1_1\propto -{x+\mathcal{i}y\over \sqrt{2}},
\end{gather}
we may define a slightly different tensor operator $u_\mathcal{m}^\dagger$ and its conjugate with components $\mathcal{m}=-1,0,1$ as (see Eqs. (3.17) of Ref. \cite{Escher}):
\begin{gather}
u_{-1}^\dagger={a_x^\dagger-\mathcal{i}a_y^\dagger\over \sqrt{2}},\qquad u_{-1}={a_x+\mathcal{i}a_y\over \sqrt{2}},\label{pd-}\\
u_0^\dagger=a_z^\dagger,\qquad\qquad\qquad u_0=a_z,\label{pd0}\\
u_1^\dagger=-{a_x^\dagger+\mathcal{i}a_y^\dagger\over \sqrt{2}},\qquad u_1=-{a_x-\mathcal{i}a_y\over \sqrt{2}}.\label{pd+}
\end{gather}
Alternatively:
\begin{gather}
a_x^\dagger={u_{-1}^\dagger-u_1^\dagger\over\sqrt{2}},\qquad
a_y^\dagger=\mathcal{i}{u_{-1}^\dagger+u_1^\dagger\over\sqrt{2}},\qquad
a_z^\dagger=u_0^\dagger.\label{au}
\end{gather}

From the mathematical point of view, the $u_\mathcal{m}^\dagger$ is a spherical tensor operator of degree $\mathcal{l}=1$ (see the Appendices A and B for the proof). The physical meaning of the $u_\mathcal{m}^\dagger$ is revealed, when acting on the vacuum eigenstate of the Hamiltonian $h_{0}$, namely on the $\ket{n_z,n_x,n_y}$= $\ket{0,0,0}$ orbital:
\begin{gather}
u_{-1}^\dagger\ket{0,0,0}={\ket{0,1,0}-\mathcal{i}\ket{0,0,1}\over \sqrt{2}},\label{ap-}\\
u_0^\dagger\ket{0,0,0}=\ket{1,0,0},\label{ap0}\\
u_1^\dagger\ket{0,0,0}=-{\ket{0,1,0}+\mathcal{i}\ket{0,0,1}\over \sqrt{2}},\label{ap+}
\end{gather}
where Eqs. (\ref{pd-})-(\ref{pd+}) and (\ref{a1}) have been used. Interestingly the right hand sides of Eqs. (\ref{ap-})-(\ref{ap+}) are equal to the spherical eigenstates $\bf\ket{n,l,m_l}$ of Eqs. (\ref{m-})-(\ref{m+}) respectively. Therefore  the operators $u_\mathcal{m}^\dagger$ create a harmonic oscillator quantum with angular momentum $l=1$ and projection of the angular momentum $m_l=\mathcal{m}=\pm 1, 0$, when acting on the vacuum state.

Since the quanta are bosons, the $u_\mathcal{m}^\dagger$ operators must obey the boson commutators. Indeed with the definitions (\ref{pd-})-(\ref{pd+}), the identities (\ref{id1}), (\ref{id3}) and the commutators of Eqs.  (\ref{b1}) one may prove that:
\begin{gather}
[u_\mathcal{m},u_\mathcal{m'}^\dagger]=\delta_{\mathcal{mm'}},\qquad
[u_\mathcal{m}^\dagger,u_\mathcal{m'}^\dagger]=[u_\mathcal{m},u_\mathcal{m'}]=0.\label{b3}
\end{gather}

The spin-orbit interaction $\mathbf{l}\cdot \mathbf{s}$ has to be added in the nuclear Hamiltonian:
\begin{equation}\label{H}
h=h_{0}+\upsilon_{ls}\hbar\omega \mathbf{l}\cdot \mathbf{s},
\end{equation}
where $\bf s$ is the spin of the particle, and $\upsilon_{ls}$ is the strength parameter of the spin-orbit interaction (see Table I of \cite{Bengtsson1985} and \cite{Nilsson2}). The spin-orbit interaction leads to the derivation of the total angular momentum:
\begin{equation}\label{lsj}
{\bf j}={\bf l}+{\bf s}.
\end{equation}
Thus the spinor $\ket{s,m_s}$ with $s={1\over 2}$ and $m_s=\pm{1\over 2}$ must also be considered. Consequently the single particle states may be written as:
\begin{gather}
{\bf\ket{n,l,m_l}}\ket{s,m_s}={\bf \ket{n,l,m_l,m_s}},\label{sphun}\\
\ket{n_z,n_x,n_y}\ket{s,m_s}=\ket{n_z,n_x,n_y,m_s},\label{cartun}
\end{gather} 
having in mind that a unitary transformation among the two bases of Eqs. (\ref{sphun}), (\ref{cartun}) exists \cite{proxy4}. 

The coupling of the spatial part of the wave function $\bf\ket{n,l,m_l}$ with the spinor $\ket{s={1\over 2},m_s=\pm{1\over 2}}$ leads to the Shell Model states:
\begin{equation}
\ket{n,l,j,m_j}=\sum_{m_l,m_s} (lm_lsm_s|jm_j){\bf\ket{n,l,m_l,m_s}},
\end{equation}
with $m_j$ being the projection of the total angular momentum and $(lm_lsm_s|jm_j)$ are the Clebsch-Gordan coefficients \cite{Edmonds}. The $\ket{n,l,j,m_j}$ denote the usual Shell Model orbitals, if one adds 1 unit in the radial quantum number $n$ and represents the angular momentum $l=0,1,2,...$ by the small latin characters $s,p,d,...$. For instance the orbital $\ket{n,l,j,m_j}$= $\ket{0,1,{3\over2},{1\over 2}}$, is labeled $1p^{j=3/2}_{m_j=1/2}$. 

The spherical states $\ket{n,l,j,m_j}$ can be transformed to the cartesian states $\ket{n_z,n_x,n_y,m_s}$ as in Ref. \cite{proxy4}:
\begin{gather}
\ket{n,l,j,m_j}=\nonumber\\
\sum_{n_z+n_x+n_y=2n+l} \braket{n_z,n_x,n_y,m_s|n,l,j,m_j}\ket{n_z,n_x,n_y,m_s}
\end{gather}
Consequently one may consider the $\ket{n_z,n_x,n_y,m_s}$ states as an {\it alternative Shell Model basis}, expressed in the cartesian coordinate system. The necessity for this cartesian basis is demonstrated by Elliott and Harvey in Refs. \cite{Elliott3,Harvey}.

\section{The Shell Model SU(3) symmetry}\label{Elliott}

A very instructive illustration of the algebraic chains, which lead from the valence Shell Model space to the Shell Model $SU(3)$ symmetry lies in the Figure 7.1 of Ref. \cite{Bible}. We shall discuss the algebraic chains and their physical meaning in this article for completeness.

The 3D-HO Hamiltonian of Eq. (\ref{h0}) has eigenstates, which constitute the harmonic oscillator shells. The eigenstates of the $h_{0}$ of the harmonic oscillator shell with $\mathcal{N}=$0, 1, 2, 3, 4, 5, 6 quanta lie among the proton or neutron magic numbers 0-2, 2-8, 8-20, 20-40, 40-70, 70-112, 112-168 respectively.

Such harmonic oscillator shells, which consist by orbitals with common number of quanta $\mathcal{N}$, posses the:
\begin{gather}
U(4\Omega)=U(\Omega)\otimes U(4)
\end{gather}
symmetry \cite{Elliott1,Elliott2}, where $\Omega={(\mathcal{N}+1)(\mathcal{N}+2)\over 2}$ is the number of the spatial harmonic oscillator eigenstates (for instance the $\ket{n_z,n_x,n_y}$ or the $\bf\ket{n,l,m_l}$) and $4$ stands for the four possible projections of spin and isospin $m_s=\pm{1\over2}, m_t=\pm {1\over 2}$, a nucleon may adopt. As an example the shell with $\mathcal{N}=0$, lies among the magic numbers 0-2, contains 1 orbital $\ket{0,0,0}$, accommodates up to 2 protons and 2 neutrons and possesses a $U(4)=U(1)\otimes U(4)$ symmetry. This $U(4\Omega)$ algebra has totally antisymmetric irreps. The $U(4)$ symmetry of the isospin is called the Wigner $SU(4)$ symmetry \cite{Wigner1937}.

In order to make the concept of the Shell Model $SU(3)$ symmetry clear, we will work out an example from the basic Quantum Mechanics through out the text. In our example we will suppose that a nucleus has 2 valence protons in the s, d nuclear shell, which lies among the proton magic numbers 8-20. This valence shell consists by orbitals with $\mathcal{N}=2$ number of quanta. Thus the 2 protons shall occupy the cartesian orbital $\ket{n_z,n_x,n_y,m_s,m_t}$ = $\ket{2,0,0,\pm{1\over 2},+{1\over 2}}$, according to the highest weight irrep (see Refs. \cite{Elliott2, Martinou2021,proxy2,proxy5,Bonatsos2020} for the explanation). In this article $m_t=+{1\over 2}$ for protons, while $m_t=-{1\over 2}$ for neutrons. The wave function of the 2 protons has to be totally antisymmetric Slater determinant \cite{Slater}, according to the Pauli Principle \cite{Pauli,Fermi}. If the state:
\begin{gather}
\phi^{m_s,m_t}(i_1)=\ket{n_z,n_x,n_y,m_s,m_t}_{i_1}=\nonumber\\
\phi(i_1)\ket{m_s,m_t}_{i_1}
\end{gather}
represents the orbital of the $i_1^{th}$ nucleon, with the spin-isospin part being: 
\begin{gather}
\ket{s,m_s}_{i_1}\ket{t,m_t}_{i_1}=\ket{m_s,m_t}_{i_1}=\ket{\pm, \pm}_{i_1}
\end{gather}
while the spatial part is:
\begin{gather}
\phi(i_1)=\ket{n_z,n_x,n_y}_{i_1}
\end{gather}
then the wave function of the two particles is the Slater determinant:
\begin{gather}\label{Slater}
\Phi={1\over\sqrt{2!}} 
\begin{vmatrix}
\phi^{++}(1) & \phi^{-+}(1) \\ 
\phi^{++}(2) & \phi^{-+}(2) \\
\end{vmatrix}.
\end{gather}
The $\phi^{m_s,m_t}$ are the states of the $4\Omega$ space, with $\Omega=6$ for the s, d nuclear shell. The irreps of the $U(4\Omega)$ symmetry show the ways one may place the $A_{val}$ objects (valence protons and neutrons) in the $4\Omega$ states.

Then the $U(4)$ symmetry is decomposed into the nuclear spin ($S$) and the nuclear isospin ($T$) symmetries:
\begin{gather}
U(\Omega)\otimes U(4)\rightarrow U(\Omega)\otimes [SU_S(2)\otimes SU_T(2)]
\end{gather}
Emphasis has to be given in the fact that the spatial part of the wave function, which is represented by the $U(\Omega)$ algebra, is treated {\it separately} by the spin and the isospin part, which are represented by the $SU_S(2)$ and the $SU_T(2)$ algebras respectively.

To make this statement clear, we will go on with our example. If the $LS$ coupling scheme is to be followed, {\it i.e.},:
\begin{gather}
{\bf L}=\sum_i{\bf l}(i),\qquad {\bf S}=\sum_i{\bf s}(i),\qquad {\bf J}={\bf L}+{\bf S}\label{LS}
\end{gather}
(with ${\bf l}(i),{\bf s}(i)$ being the angular momentum and the spin respectively of the $i^{th}$ particle), then the Slater determinant of Eq. (\ref{Slater}) can be decomposed into a spatial part and a spin-isospin part:
\begin{gather}
\Phi=\left(\phi(1)\phi(2)\right)\nonumber\\
\left({1\over\sqrt{2!}}(\ket{++}_1\ket{-+}_2-\ket{++}_2\ket{-+}_1)\right).
\end{gather}
Obviously the spatial part of the wave function is:
\begin{gather}
\Phi_{space}=\phi(1)\phi(2)
\end{gather}
and it is totally symmetric in the transposition of the two particles, while the spin-isospin part is:
\begin{gather}
\Phi_{spin-isospin}={1\over\sqrt{2!}}(\ket{++}_1\ket{-+}_2-\ket{++}_2\ket{-+}_1)
\end{gather}
and it is totally antisymmetric in the transposition of the particles. The overall product of the wave functions \begin{gather}
\Phi=\Phi_{space}\cdot\Phi_{spin-isospin}
\end{gather}
is thus antisymmetric, as it should be according to the Pauli Principle. The spatial part of the wave function possesses the $U(\Omega)$ symmetry, while the spin-isospin the $U(4)$.

This is the very essence of the $LS$ coupling scheme: that the spatial part of the nuclear wave function generates the nuclear angular momentum $L$, the spinor part generates the nuclear spin $S$ and that one may treat these two parts {\it separately}, as long as the product of the two of them respects the Pauli Principle for the multi fermion system. The antisymmetry of the overall multi nucleon wave function is guaranteed if the Young diagram of spin-isospin part ($U(4)$) is the conjugate of the Young diagram of the spatial part ($U(\Omega)$). The interested reader can find more details about this conjugation in Section 7.1.2 of Ref. \cite{Bible} of Draayer's chapter or in Chapter 29 of Talmi's book \cite{Talmi}. This separation of the spatial wave function from the spin-isospin part is achieved in the $LS$ coupling scheme and leads to the Shell Model $SU(3)$ symmetry.

A significant spin-orbit splitting of the single-nucleon energies may cause the rise of the spin-orbit like shells, among proton or neutron numbers 6-14, 14-28, 28-50, 50-82, 82-126,126-182 \cite{Haxel1949}. These shells consist by harmonic oscillator eigenstates with $\mathcal{N}$ and $\mathcal{N}+1$ quanta (see Table 7 of Ref. \cite{proxy4}) and so the $U(4\Omega)=U(\Omega)\otimes U(4)$ symmetry no longer has a straightforward application. 

One of the possible ways \cite{Cseh2018,Kota} to overpass this problem is the use of the proxy-$SU(3)$ symmetry \cite{proxy1,Cakirli2006,Sofia2013}. In this type of approximate symmetry, one may apply a unitary transformation in the intruder orbitals with $\mathcal{N}+1$ quanta \cite{proxy4}, so as to transform them to their de Shalit-Goldhaber counterparts \cite{deShalit}. This unitary transformation \cite{Bonatsos2021} reduces the total number of quanta of the intruder orbitals by 1 unit ($\mathcal{N}+1\rightarrow \mathcal{N}$) and it is similar in spirit with the unitary transformation introduced in the pseudo $SU(3)$ symmetry \cite{AnnArbor,Draayer1984,Castanos1987}. 

The advantages of the proxy-$SU(3)$ symmetry are the following:\\
a) the relation of the intruder orbitals to their proxies is based on the experimental observations of de Shalit and Goldhaber \cite{deShalit} and of Cakirli, Blaum and Casten \cite{Burcu2010},\\
b) the unitary transformation used in the proxy-$SU(3)$ symmetry leaves the normal parity orbitals (those with $\mathcal{N}$ quanta) intact, and affects only the intruder orbitals (those with $\mathcal{N}+1$ quanta),\\
c) the proxy transformation affects only the $z$ axis of the intruder orbitals and so the the number of quanta in the $x,y$ plane is conserved. This means that the projection of the total and the orbital single particle angular momenta, which are good quantum numbers in the deformed nuclei \cite{Bonatsos2020a,Sobhani2021}, are not affected by the transformation. We have zero error in the prediction of the band label $K$ and minimum error in the cut off the nuclear angular momentum ($L_{max}$) for each band \cite{proxy4}.\\
Furthermore the irreps of the proxy-$SU(3)$ symmetry have given parameter free predictions for the prolate-oblate transition \cite{proxy2} and for the islands of inversion and shape coexistence \cite{Martinou2021,Martinou2021a}, while within a single parameter they give promising early stage results for the binding and the two neutron separation energies \cite{Martinou2021b}.

As a result, in a harmonic oscillator shell, one may use the $U(4\Omega)=U(\Omega)\otimes U(4)$ symmetry in a straightforward way as in Refs. \cite{Elliott1,Elliott2,Elliott3,Elliott4}, while in a spin-orbit like shell the $U(4\Omega)=U(\Omega)\otimes U(4)$ can be approximately applied within the proxy-$SU(3)$ scheme \cite{proxy2,proxy3,proxy4}. The gain is that in any of the two types of shells, the spatial $U(\Omega)$ symmetry exists and is decomposed as \cite{Elliott1,Elliott2}:
\begin{gather}
 U(\Omega)\supset U(3)\supset SU(3)\supset O(3)\supset O(2),\label{chain1}
\end{gather}
Clearly the Shell Model $SU(3)$ symmetry derives from the spatial $U(\Omega)$ symmetry. The labels of each of the above symmetries are \cite{Lipas}:
\begin{gather}
[f]=[f_1,f_2,...,f_{\Omega}]:\mbox{ for the }U(\Omega),\nonumber\\
[f_1,f_2,f_3]:\mbox{ for the }U(3),\nonumber\\
(\lambda,\mu):\mbox{ for the } SU(3),\nonumber\\
L:\mbox{ for the }O(3), \nonumber\\
M:\mbox{ for the }O(2),
\end{gather}
where $M$ is the projection of the nuclear orbital angular momentum.

In our example, the one of the two protons in the s, d shell the irrep of the $U(\Omega=6)$ is $[2,0,0,0,0,0]$ since the two protons occupy the same $\ket{n_z,n_x,n_y}$ orbital, the irrep of the $U(3)$ is $[4,0,0]$ since in the highest weight irrep $f_1=\sum_{i}n_{z}(i)$, $f_2=\sum_in_{x}(i)$ and $f_3=\sum_in_{y}(i)$ \cite{Elliott2,proxy5} and $(\lambda,\mu)=(4,0)$ since $\lambda=f_1-f_2$ and $\mu=f_2-f_3$. The subscript $i$ is for every valence nucleon. Therefore the spatial part of the Shell Model $SU(3)$ wave function is labeled as:
\begin{gather}
\Phi_{spatial}([f](\lambda,\mu)).
\end{gather}
The spin-isospin part, which is the conjugate of the spatial, is labeled by the nuclear spin ${\bf S}$, the nuclear isospin ${\bf T}=\sum_i {\bf t}(i)$ and their projections $M_S,M_T$:
\begin{gather}
\Phi_{spin-isospin}(T,M_T,S,M_S)
\end{gather}
In our example $S=0$, $M_S=0$, $T=1$, $M_T=1$. Thus the overall nuclear wave function is labeled by the \cite{Elliott3}:
\begin{gather}
\Phi(TS[f](\lambda,\mu)M_TM_S).
\end{gather}
Despite the fact that the overall Shell Model $SU(3)$ wave function is labeled by both the spatial and the spin-isospin irreps, one has to remember that the $U(3)$ and $SU(3)$ lie solely in the spatial part of the state and this is the privilege of the $LS$ coupling scheme.

The Shell Model $U(3)$ algebra is generated by the 9 cartesian generators of the form \cite{Elliott3,Harvey}:
\begin{gather}
C_{k,k'}=a^\dagger_ka_{k'},\mbox{ with }k,k'=x,y,z.\label{gen1}
\end{gather}
The 3 components of the angular momentum $L_z,L_\pm$, the 5 components of the quadrupole operator $Q_\mathcal{m},\mathcal{m}=\pm 2, \pm 1,0$ and the number (of quanta) operator can be expressed as linear combinations of the generators of Eq. (\ref{gen1}) and their commutators close the $U(3)$ algebra \cite{Elliott1,Elliott2}.

Taking advantage of the equivalence of the $u_\mathcal{m}^\dagger,u_\mathcal{m}$ with the $a_k^\dagger,a_k$ operators, which derives from the Eqs. (\ref{pd-})-(\ref{pd+}), one may construct the spatial $U(3)$ algebra of a valence shell from the {\it spherical quanta states}. In this scenario the quanta are created by spherical tensors of degree $\mathcal{l}=1$ and thus they may be arranged according to the 3 components $\mathcal{m}=\pm 1,0$, instead of being arranged according to the 3 cartesian directions of the Elliott-Harvey point of view \cite{Elliott3,Harvey}. The $U(3)$ algebra of the spherical quanta is generated by the 9 operators of the form:
\begin{gather}\label{gen2}
\mathcal{A_{m,m'}}=u_\mathcal{m}^\dagger u_\mathcal{m'},\mbox{ with }\mathcal{m,m'}=\pm 1,0.
\end{gather}
Using the boson commutators (\ref{b3}) along with the identities (\ref{id5}), (\ref{id6}) one may calculate all the commutators of the type $[\mathcal{A_{m,m'}},\mathcal{A_{m'',m'''}}]$ with $\mathcal{m,m',m'',m'''}=\pm 1,0$ and produce the Multiplication Table  (see Table \ref{U(3)}). Since the set of generators of an algebra is not unique, one may consider the $C_{k,k'}$ of the expression (\ref{gen1}) and the $\mathcal{A_{m,m'}}$ of (\ref{gen2}) as two generator sets of the Shell Model $U(3)$ algebra.

\begin{table*}
\caption{Multiplication table of the $U(3)$ algebra, which is generated by the operators $\mathcal{A_{m,m'}}$. For instance $[\mathcal{A_{1,0}},\mathcal{A_{1,1}}]$= $ - \mathcal{A_{1,0}}$.  }\label{U(3)}
\begin{tabular}{c|c|c|c|c|c|c|c|c|c}
 $\mathcal{A_{m,m'}}$ & $\mathcal{A_{1,1}}$ &  $\mathcal{A_{1,0}}$ &  $\mathcal{A_{1,-1}}$ &  $\mathcal{A_{0,1}}$ &  $\mathcal{A_{0,0}}$ &  $\mathcal{A_{0,-1}}$ &  $\mathcal{A_{-1,1}}$ &  $\mathcal{A_{-1,0}}$ &  $\mathcal{A_{-1,-1}}$\\
\hline
$\mathcal{A_{1,1}}$ & 0 &  $\mathcal{A_{1,0}}$ &  $\mathcal{A_{1,-1}}$ & - $\mathcal{A_{0,1}}$ & 0 & 0 & - $\mathcal{A_{-1,1}}$ & 0 & 0\\
 $\mathcal{A_{1,0}}$ & -$\mathcal{A_{1,0}}$ & 0 & 0 &  $\mathcal{A_{1,1}}$-$\mathcal{A_{0,0}}$ &  $\mathcal{A_{1,0}}$ & $\mathcal{A_{1,-1}}$ & -$\mathcal{A_{-1,0}}$ & 0 & 0\\
 $\mathcal{A_{1,-1}}$ & -$\mathcal{A_{1,-1}}$ & 0 & 0 & -$\mathcal{A_{0,-1}}$ & 0 & 0 &  $\mathcal{A_{1,1}}$-$\mathcal{A_{-1,-1}}$ & $\mathcal{A_{1,0}}$ & $\mathcal{A_{1,-1}}$\\
$\mathcal{A_{0,1}}$ & $\mathcal{A_{0,1}}$ &  $\mathcal{A_{0,0}}$- $\mathcal{A_{1,1}}$ &  $\mathcal{A_{0,-1}}$ & 0 & - $\mathcal{A_{0,1}}$ & 0 & 0 & - $\mathcal{A_{-1,1}}$ & 0\\
 $\mathcal{A_{0,0}}$ & 0 & - $\mathcal{A_{1,0}}$ & 0 &  $\mathcal{A_{0,1}}$ & 0 &  $\mathcal{A_{0,-1}}$ & 0 & -$\mathcal{A_{-1,0}}$ & 0\\ 
 $\mathcal{A_{0,-1}}$ & 0 & - $\mathcal{A_{1,-1}}$ & 0 & 0 &- $\mathcal{A_{0,-1}}$ & 0 &  $\mathcal{A_{0,1}}$ &  $\mathcal{A_{0,0}}$- $\mathcal{A_{-1,-1}}$ &  $\mathcal{A_{0,-1}}$\\
 $\mathcal{A_{-1,1}}$ &  $\mathcal{A_{-1,1}}$ &  $\mathcal{A_{-1,0}}$ &  $\mathcal{A_{-1,-1}}$- $\mathcal{A_{1,1}}$ & 0 & 0 & - $\mathcal{A_{0,1}}$ & 0 & 0 & - $\mathcal{A_{-1,1}}$\\ 
 $\mathcal{A_{-1,0}}$ &  0 & 0 & - $\mathcal{A_{1,0}}$ &  $\mathcal{A_{-1,1}}$ &  $\mathcal{A_{-1,0}}$ &  $\mathcal{A_{-1,-1}}$- $\mathcal{A_{0,0}}$ & 0 & 0 & - $\mathcal{A_{-1,0}}$\\ 
  $\mathcal{A_{-1,-1}}$ & 0 & 0 & - $\mathcal{A_{1,-1}}$ & 0 & 0 & - $\mathcal{A_{0,-1}}$ &  $\mathcal{A_{-1,1}}$ &  $\mathcal{A_{-1,0}}$ & 0\\
\end{tabular}
\end{table*}

\section{The Shell Model SU(3) states\label{notation}}

When one is working in the level of the $U(\Omega)$ symmetry, the irreps $[f_1,f_2,...,f_{\Omega}]$ represent with how many and with which ways the ``objects'' can be placed in the $\Omega$ ``states''. In this level the ``objects'' are the indistinguishable valence nucleons and the ``states'' are the spatial orbitals $\ket{n_z,n_x,n_y}$. Draayer, Leschber, Park and Lopez in Ref. \cite{code} accomplished the $U(\Omega)\supset U(3)$ decomposition. This is a pure mathematical procedure, but what is the physical meaning of this decomposition? 

The fact is that when one is working in the level of the $U(3)$ symmetry, the irreps $[f_1,f_2,f_3]$ represent in how many many ways one may place the ``objects'' in a three dimensional  space. Now the three dimensions are the Hermite polynomials $\ket{n_z=1},\ket{n_x=1},\ket{n_y=1}$, which are eigenstates of the harmonic oscillator, while the ``objects'' are the indistinguishable harmonic oscillator quanta, which derive from the placement of the nucleons in the $\ket{n_z,n_x,n_y}$ states. For instance the Shell Model $U(3)$ irrep $[f_1,f_2,f_3]=[2,1,0]$ is about two quanta, which have occupied the state $\ket{n_z=1}$ and about one quantum in the state $\ket{n_x=1}$. Therefore the objects of the $U(3)$ wave functions are not anymore the nucleons, but the quanta. Thus the many nucleon wave functions of the $U(\Omega)$ symmetry, are being decomposed to the many quanta wave functions of the $U(3)$ symmetry. This is the very meaning of the decomposition Draayer {\it et al.} accomplished in Ref. \cite{code}. The $U(3)$ irreps $[f_1,f_2.f_3]$ show with many and with which ways one may transpose the harmonic oscillator quanta in the three cartesian axes. Each transposition of the harmonic oscillator quanta is equivalent with a spatial rotation \cite{Troltenier1996}.

The third article of the Shell Model $SU(3)$ symmetry was written by Elliott and Harvey. Harvey wrote another article (see Ref. \cite{Harvey}) where he explained the details of the model. We shall now focus in Eq. (3.15) of section 3.3 of Harvey's article in Ref. \cite{Harvey}. There he presented that the $U(3)$ wave function is made of states:
\begin{gather}
\ket{pqr}_{i_1,...,i_{p+q+r}}=
a_z^\dagger(i_1)a_z^\dagger(i_2)...a_z^\dagger(i_p)a_x^\dagger(i_{p+1})...a_x^\dagger(i_{p+q})\nonumber\\
a_y^\dagger(i_{p+q+1})...a_y^\dagger(i_{p+q+r})\ket{0}.\label{U3state}
\end{gather}
The dagger operators are those introduced in Eq. (\ref{a}). The labels $i_1,...,i_{p+q+r}$ take the values $1,2,3,...,A_{val}$, where $A_{val}$ is the valence number of nucleons. It is possible that a particle number may appear more than once, or not at all; so it is possible that $i_1=i_2=1$. The $a_z^\dagger (i_1)\ket{0}$ represents a quantum in the $z$ axis from the $i_1^{th}$ particle. Clearly in the state of Eq. (\ref{U3state}) there are $p$ quanta in the $z$ axis, $q$ quanta in the $x$ axis and $r$ quanta in the $y$ axis. So the numbers $1,2,...,p,p+1,...,p+q,p+q+1,...,p+q+r$ enumerate the quanta, which are the ``objects'' of the $U(3)$ symmetry. Indeed the quanta are being enumerated and this is necessary for the construction the particle-number Young tableau of the $U(3)$ states, which will be discussed afterwards. 

Harvey wrote that the vacuum $\ket{0}$ is the state of no quanta, namely the $1s$ orbital. We have already emphasized that the $U(3)$ symmetry derives solely from the spatial part of the many nucleon wave function (not from the overall, not from the spin-isospin part) and so the vacuum could not be the $1s^{j=1/2}$ orbital; the $j$ quantum number could not be included in the vacuum state, when one is building the $U(3)$ states. As already outlined, in the $U(3)$ states the ``objects'' are the quanta and there exist $p+q+r$ of them, thus the vacuum state in Eq. (\ref{U3state}) is:
\begin{gather}
\ket{0}=\ket{0(i_1),0(i_2),...,0(i_{p+q+r})},
\end{gather}
where by the $\ket{0(i_1)}$ we mean that there are no quanta due to the $i_1^{th}$ particle etc.

For instance the action of one dagger operator is:
\begin{gather}
a_z(i_1)^\dagger\ket{0(i_1),0(i_2),...,0(i_{p+q+r})}=\nonumber\\
\ket{1_z(i_1),0(i_2),...,0(i_{p+q+r})},
\end{gather}
where $\ket{1_z(i_1)}$ represents one quantum in the $z$ axis deriving from the $i_1^{th}$ particle, namely the Hermite polynomial $H_{1}(bz_{i_1})=\ket{1_z(i_1)}$ with $b=\sqrt{m\omega\over\hbar}$ (see Complement $B_{VII}$ of Ref. \cite{Cohen}). Accordingly the action of two dagger operators is:
\begin{gather}
a_z^\dagger(i_1) a_z^\dagger(i_2)\ket{0(i_1),0(i_2),...,0(i_{p+q+r})}=\nonumber\\
\ket{1_z(i_1),1_z(i_2),...,0(i_{p+q+r})},\label{az2}
\end{gather}
and this represents 1 quantum in the $z$ axis deriving from the $i_1^{th}$ particle and 1 more deriving from the $i_2^{th}$ particle.

In order to keep working on our example, the one of the 2 protons in the s, d shell, I shall present the $U(3)$ state of it:
\begin{gather}
a_z^\dagger(i_1)a_z^\dagger(i_2)a_z^\dagger(i_3)a_z^\dagger(i_4)\ket{0}=\nonumber\\
\ket{1_z(i_1),1_z(i_2),1_z(i_3),1_z(i_4)},
\end{gather}
where $i_1=i_2=1$, $i_3=i_4=2$. So we have 2 quanta in the $z$ axis due to the first proton and two more quanta in the $z$ axis due to the second proton.

The many quanta $U(3)$ wave function of Eq. (\ref{U3state}) can be represented by a Young tableau. Each box in a Young tableau is an ``object'', which in the case of the Shell Model $U(3)$ symmetry is a harmonic oscillator quantum in one cartesian axis. A general quantum-number (left) and particle-number (right) Young tableau \cite{Lipas} of the Shell Model $U(3)$ symmetry looks like the:
\begin{equation}\label{123}
\begin{Young}
{\bf z}&{\bf z}&...&...&{\bf z} \cr
{\bf x}&...&...&{\bf x}\cr
{\bf y}&...&{\bf y}\cr\end{Young}\qquad
\begin{Young}
1&2&3&...&...\cr
...&...&...&...\cr
...&...&...\cr\end{Young}
\end{equation}
The labels $\bf z,x,y$ on the left signify a quantum in the $z,x,y$ cartesian axis respectively. The numbers on the right enumerate the quanta, take values $1,2,...,p,p+1,...,p+q,p+q+1,...,p+q+r$ using Harvey's notation (Eq. (3.15) of Ref. \cite{Harvey}) and can be placed in the boxes so as to increase from the left to the right and from up to down \cite{Lipas}. The position of the numbers indicates the permutation symmetry of the quanta. The permutation of the quanta is discussed extensively in Ref. \cite{proxy5}.

It is common practice, that in a $U(3)$ Young tableau a column with three boxes is erased. In this way one reproduces the Shell Model $SU(3)$ Young tableaux of the highest weight irrep:
\begin{equation}\label{123SU(3)}
\begin{Young}
{\bf z}&{\bf z}&...&...&{\bf z} \cr
{\bf x}&...&...&{\bf x}\cr
\end{Young}\qquad
\begin{Young}
1&2&3&4&...\cr
...&...&...&...\cr
\end{Young}
\end{equation}
In general two boxes in a row of a Young diagram
\begin{equation}\label{123}
\begin{Young}{\bf z}&{\bf z} \cr\end{Young}\qquad\begin{Young}1&2\cr\end{Young}
\end{equation}
represent a symmetric pair of quanta, while two boxes in a column
\begin{equation}\label{123}
\begin{Young}{\bf z}\cr {\bf x} \cr\end{Young}\qquad\begin{Young}1\cr 2\cr\end{Young}
\end{equation}
correspond to an antisymmetric pair of quanta. At this point recall that since the quanta are bosons they can form symmetric and antisymmetric pairs, in contradiction with the fermions, which can form only antisymmetric pairs. From the above becomes clear that the Shell Model $U(3)$ symmetry has to do with harmonic oscillator quanta, which are {\bf bosons} and are coupled into symmetric or into antisymmetric pairs. 

The Shell Model $SU(3)$ labels for the highest weight irrep are \cite{Elliott2,proxy5}:
\begin{gather}
\lambda=f_1-f_2=\sum_{i}n_{z}(i)-\sum_{i}n_{x}(i),\\
\mu=f_2-f_3=\sum_{i}n_{x}(i)-\sum_{i}n_{y}(i).
\end{gather}
A general irrep $(\lambda,\mu)$ with $\mu\ne 0$ represents an $SU(3)$ state of mixed symmetry, {\it i.e.}, it is not totally symmetric \cite{proxy5}.

A symmetric state of two quanta in the $k,k'$ cartesian directions may be labeled:
\begin{gather}
a_k^\dagger (i_w) a_{k'}^\dagger(i_{w'})\ket{0}=\nonumber\\
{1\over\sqrt{2!}}\Big(\ket{1_k(i_w),1_{k'}(i_{w'})}+\ket{1_k(i_{w'}),1_{k'}(i_w)}\Big),\label{block}
\end{gather}
where $w\ne w'$ and they obtain the values $1,2,...,p+q+r$, while the $k,k'$ can be the $z,x,y$. This state is symmetric in the permutation $w\leftrightarrow w'$. Since this state represents a symmetric pair of quanta in the $kk'$ axes, I shall call the normalized state of one pair of symmetric harmonic oscillator quanta:
\begin{gather}
|n_{kk'}=1)={1\over\sqrt{2!}}\Big(\ket{1_k(i_w),1_{k'}(i_{w'})}+\ket{1_k(i_{w'}),1_{k'}(i_w)}\Big),
\end{gather}
where $n_{kk'}$ counts the number of the pairs of the symmetric quanta in the $kk'$ axes. Note that the ket of this state is round $|)$. The vacuum of this space shall be the state of no pairs of symmetric quanta:
\begin{gather}
|0)=|n_{kk'}=0).
\end{gather}

The action of the $a_k^\dagger (i_1) a_{k'}^\dagger(i_2)$ on the vacuum creates a symmetric pair of quanta in the $kk'$ axes:
\begin{gather}
a_k^\dagger (i_1) a_{k'}^\dagger(i_2)|0)=|1).
\end{gather}
The action of this operator once more on the vacuum shall create a state of two symmetric pairs of quanta:
\begin{gather}
c_1\left(a_k^\dagger (i_1) a_{k'}^\dagger(i_2)\right)\left(a_k^\dagger (i_3) a_{k'}^\dagger(i_4)\right)|0)=|2),
\end{gather}
where $c_1$ is a normalization constant. The above state is the:
\begin{gather}
|2)=c_1\left(a_k^\dagger (i_1) a_{k'}^\dagger(i_2)\ket{0}\right)\left(a_k^\dagger (i_3) a_{k'}^\dagger(i_4)\ket{0}\right)\nonumber\\
+c_1\left(a_k^\dagger (i_3) a_{k'}^\dagger(i_4)\ket{0}\right)\left(a_k^\dagger (i_1) a_{k'}^\dagger(i_2)\ket{0}\right).\label{Fock2}
\end{gather}
Since there are $2!$ terms in the above summation, the normalization constant has to be $c_1={1\over\sqrt{2!}}$. This state is symmetric in the permutation of the pairs $(1,2)\leftrightarrow(3,4)$. 

The normalized state of three symmetric pairs of quanta shall be:
\begin{gather}
c_2\left(a_k^\dagger (i_1) a_{k'}^\dagger(i_2)\right)\left(a_k^\dagger (i_3) a_{k'}^\dagger(i_4)\right)\left(a_k^\dagger (i_5) a_{k'}^\dagger(i_6) \right)|0)=\nonumber\\
|3),\label{Fock3}
\end{gather}
where $c_2$ is a normalization constant (see Eq. (2.2) of Ref. \cite{Lipas} for the expression). The state $|3)$ consists by a summation of $3!$ terms, just like the state $|2)$ consisted by $2!$ terms. So the normalization constant shall be $c_2={1\over\sqrt{3!}}$. The $|3)$ is symmetric in the permutation of the pairs of quanta $(1,2)\leftrightarrow (3,4)\leftrightarrow (5,6)$.

Consequently a symmetric and normalized state of $N_{kk'}$ symmetric pairs of quanta in the $kk'$ axes can be written:
\begin{gather}
|n_{kk'}=N_{kk'})=\nonumber\\
{1\over\sqrt{N_{kk'}!}}\left(a_k^\dagger(i_{2n_{kk'}+1}) a_{k'}^\dagger(i_{2n_{kk'}+2})\right)^{N_{kk'}}|n_{kk'}=0).\label{Fock}
\end{gather}
From Eq. (\ref{Fock}) we may find the action of the pair (of quanta) creation operator on the $|n_{kk'})$ state:
\begin{gather}
a_k^\dagger(i_1) a_{k'}^\dagger(i_2)|0)=|1),\nonumber\\
{1\over\sqrt{2}}a_k^\dagger(i_3) a_{k'}^\dagger(i_4)|1)=|2)\Rightarrow a_k^\dagger(i_3) a_{k'}^\dagger(i_4)|1)=\sqrt{2}|2),\nonumber\\
{1\over\sqrt{3}}a_k^\dagger(i_5) a_{k'}^\dagger(i_6)|2)=|3)\Rightarrow a_k^\dagger(i_5) a_{k'}^\dagger(i_6)|2)=\sqrt{3}|3),\nonumber\\
 a_k^\dagger(i_{2n_{kk'}+1}) a_{k'}^\dagger(i_{2n_{kk'}+2})|n_{kk'})=\sqrt{n_{kk'}+1}|n_{kk'}+1).\label{act1}
\end{gather}

Despite that the quanta are indistinguishable particles and so we could drop the $i$ labels, I shall keep them till the reader gets used in the procedure of the construction of the $SU(3)$ wave functions. Now the annihilation operator of a symmetric pair of quanta in the $kk'$ axes shall be formed:
\begin{gather}
a_{k'}(i_{w'}) a_{k}(i_w)=\left(a_k^\dagger(i_w) a_{k'}^\dagger(i_{w'})\right)^\dagger.
\end{gather}
This operator has to destroy the {\bf same} pair of quanta, which has been created by the $a_k^\dagger(i_w) a_{k'}^\dagger(i_{w'})$ operator on the $|n_{kk'})$ state. This is achieved if:
\begin{gather}
\Big(a_{k'}(i_{2n_{kk'}})a_k(i_{2n_{kk'}-1})\Big)\Big(a_k^\dagger(i_{2n_{kk'}+1})a_{k'}^\dagger(i_{2n_{kk'}+2}) \Big)
|n_{kk'}),
\end{gather}
where Eq. (\ref{act1}) has been used.
Thus:
\begin{gather}
a_{k'}(i_{2n_{kk'}})a_k(i_{2n_{kk'}-1})=\left(a_k^\dagger(i_{2n_{kk'}+1})a_{k'}^\dagger(i_{2n_{kk'}+2}) \right)^\dagger,\label{conjugate}
\end{gather}
in order that the creation and annihilation operators create and annihilate respectively the same pair of quanta when acting on a $|n_{kk'})$ state.

Let the action of the annihilation operator be:
\begin{gather}
a_{k'}(i_{2n_{kk'}})a_k(i_{2n_{kk'}-1})|n_{kk'})=c|n_{kk'}-1),
\end{gather}
where $c$ is a number to be calculated. If the state $|n_{kk'})$ is normalized, {\it i.e.}, :
\begin{gather}
 (n_{kk'}|n_{kk'})=1,
\end{gather}
then we shall demand that the braket $(n_{kk'}\mbox{ }a_{k'}(i_{2n_{kk'}})a_k(i_{2n_{kk'}-1})|a_{k'}(i_{2n_{kk'}})a_k(i_{2n_{kk'}-1})\mbox{ }n_{kk'})$ is normalized as follows:
\begin{gather}
(n_{kk'}\mbox{ }a_{k'}(i_{2n_{kk'}})a_k(i_{2n_{kk'}-1})|a_{k'}(i_{2n_{kk'}})a_k(i_{2n_{kk'}-1})\mbox{ }n_{kk'})\nonumber\\
=n_{kk'}.
\end{gather}
Since Eq. (\ref{act1}) is valid, this normalization is satisfied if:
\begin{gather}
a_{k'}(i_{2n_{kk'}})a_k(i_{2n_{kk'}-1})|n_{kk'})=\sqrt{n_{kk'}}|n_{kk'}-1)\label{act2}
\end{gather}
and so:
\begin{gather}
\Big(a_k^\dagger(i_{2n_{kk'}+1}) a_{k'}^\dagger(i_{2n_{kk'}+2})\Big)\Big(a_{k'}(i_{2n_{kk'}})a_k(i_{2n_{kk'}-1})\Big)|n_{kk'})\nonumber\\
=n_{kk'}|n_{kk'}).\label{nkk'}
\end{gather}
The actions (\ref{act1}), (\ref{act2}) lead to action of the commutator on the states $|n_{kk'})$:
\begin{gather}
[a_{k'}(i_{2n_{kk'}}) a_{k}(i_{2n_{kk'}-1}),a_k^\dagger(i_{2n_{kk'}+1}) a_{k'}^\dagger(i_{2n_{kk'}+2})]|n_{kk'})\nonumber\\
=|n_{kk'}).
\end{gather}
Thus the commutator of the annihilation and creation operators of the symmetric pairs of quanta is:
\begin{gather}
[a_{k'}(i_{2n_{kk'}}) a_{k}(i_{2n_{kk'}-1}),a_k^\dagger(i_{2n_{kk'}+1}) a_{k'}^\dagger(i_{2n_{kk'}+2})]=1.\label{co1}
\end{gather}

Now we may go back to Eq. (\ref{Fock2}), where the $|2)$ represents a symmetric state of two pairs of quanta. This means that the pair formed by the $(1,2)$ quanta is symmetric upon interchange with the pair formed by the $(3,4)$ quanta. We can write this symmetry property as:
\begin{gather}
\left(a_k^\dagger (i_1) a_{k'}^\dagger(i_2)\right)\left(a_k^\dagger (i_3) a_{k'}^\dagger(i_4)\right)|0)=\nonumber\\
\left(a_k^\dagger (i_3) a_{k'}^\dagger(i_4)\right)\left(a_k^\dagger (i_1) a_{k'}^\dagger(i_2)\right)|0)\Leftrightarrow\nonumber\\
[a_k^\dagger (i_1) a_{k'}^\dagger(i_2),a_k^\dagger (i_3) a_{k'}^\dagger(i_4)]|0)=0. 
\end{gather}
The generalization of the above symmetry property reads:
\begin{gather}
[a_k^\dagger (i_{2n_{kk'}+1}) a_{k'}^\dagger(i_{2n_{kk'}+2}),a_k^\dagger (i_{2n_{kk'}+1}) a_{k'}^\dagger(i_{2n_{kk'}+2})]|n_{kk'})=\nonumber\\
0,
\end{gather}
which means that:
\begin{gather}
[a_k^\dagger (i_{2n_{kk'}+1}) a_{k'}^\dagger(i_{2n_{kk'}+2}),a_k^\dagger (i_{2n_{kk'}+1}) a_{k'}^\dagger(i_{2n_{kk'}+2})]\nonumber\\
=0. \label{co2}
\end{gather}
The conjugate of the above is:
\begin{gather}
[a_{k'} (i_{2n_{kk'}}) a_k(i_{2n_{kk'}-1}),a_{k'}(i_{2n_{kk'}}) a_k(i_{2n_{kk'}-1})] \nonumber\\
=0. \label{co3}
\end{gather}
The commutators (\ref{co1}), (\ref{co2}), (\ref{co3}) signify that the symmetric pairs of harmonic oscillator quanta satisfy the boson commutation relations.

Now we may return to our example, the one of the two protons in the s, d shell. The quantum-number and particle-number Young tableaux of this example are:
\begin{equation}\label{14}
\begin{Young}
{\bf z}&{\bf z}&{\bf z}&{\bf z} \cr
\end{Young}\qquad
\begin{Young}
1&2&3&4\cr
\end{Young}.
\end{equation}
This state has a $U(3)$ irrep $[f_1,f_2,f_3]=[4,0,0]$, or an $SU(3)$ irrep $(\lambda,\mu)=(4,0)$ \cite{Elliott2}.

The question now is, if we can construct the above state by symmetric pairs of quanta? Two symmetric pairs of quanta in the $z$ axis are represented by the Young tableaux:
\begin{equation}\label{12}
\begin{Young}
{\bf z}&{\bf z}\cr
\end{Young}\qquad
\begin{Young}
1&2\cr
\end{Young}\qquad,\qquad
\begin{Young}
{\bf z}&{\bf z}\cr
\end{Young}\qquad
\begin{Young}
3&4\cr
\end{Young}.
\end{equation}
Each of the two Young tableaux have a $U(3)$ irrep $[f_1,f_2,f_3]$ = $[2,0,0]$, or $SU(3)$ irrep $(\lambda,\mu)=(2,0)$ \cite{Elliott2}. The above two Young tableaux can be coupled (outer product) into a new Young tableau. The rules for the coupling are described in Refs. \cite{Coleman1964,Troltenier1996} and can be accomplished by the online code of Ref. \cite{Alex2011}, or even by the code of Ref. \cite{Dytrych2021}, which has far more reaching capabilities than this task. The results of this outer product are:
\begin{equation}
\begin{Young}
{\bf z}&{\bf z}\cr
\end{Young}\qquad
\begin{Young}
1&2\cr
\end{Young}\qquad\otimes\qquad
\begin{Young}
{\bf z}&{\bf z}\cr
\end{Young}\qquad
\begin{Young}
3&4\cr
\end{Young}\qquad =\nonumber
\end{equation}
\begin{equation}
\begin{Young}
{\bf z}&{\bf z}&{\bf z}&{\bf z} \cr
\end{Young}\qquad
\begin{Young}
1&2&3&4\cr
\end{Young}\qquad\oplus\nonumber
\end{equation}
\begin{equation}
\begin{Young}
{\bf z}&{\bf z} \cr
{\bf x}&{\bf x}\cr
\end{Young}\qquad
\begin{Young}
1&2\cr
3&4\cr
\end{Young}\qquad\oplus\nonumber
\end{equation}
\begin{equation}
\begin{Young}
{\bf z}&{\bf z}&{\bf z} \cr
{\bf x}\cr
\end{Young}\qquad
\begin{Young}
1&2&3\cr
4\cr
\end{Young}.
\end{equation}
The first quantum-number and particle-number Young tableau in the r.h.s. of the above equation is the fully symmetric state of the four quanta, while the rest two occurrences correspond to spatial rotations (Fig. 3 of Ref. \cite{Troltenier1996}). Consequently the fully symmetric state of (\ref{14}) may result from the symmetric coupling of two pairs of symmetric quanta.

Notice that all this discussion was about the coupling of the cartesian harmonic oscillator quanta. We ended to define a new space, in which the number of symmetric states of the symmetric pairs quanta are counted. The commutators of the creation and annihilation operators of the symmetric pairs of quanta, obey to the boson commutation relations, which in simple words means that in a Shell Model $SU(3)$ state:\\
a) we may have infinite number of quanta in a cartesian axis and\\
b) we may create infinite number of symmetric states of symmetric pairs of the harmonic oscillator quanta.\\
This conclusion is in accordance with the knowledge that the number of quanta in one cartesian axis in the $SU(3)$ wave functions is $\lambda+\mu$ \cite{Elliott2} and it is long known that it is possible that $\lambda+\mu\rightarrow \infty$.

The procedure we followed in the section is the well known technique used in Quantum Mechanics to prove the commutation relations of the bosons. To do so, we have step on the following conclusions: a) the $SU(3)$ states include only the spatial degrees of freedom so the spin and the total angular momentum must not be considered in the construction of the $SU(3)$ states b) the ``objects'' of the $SU(3)$ states are the quanta, which are bosons and so infinite of them can occupy the same state, c) the the $SU(3)$ wave functions can be constructed either by coupling the quanta, or by coupling the symmetric pairs of quanta, if the $\lambda+\mu,\mu$ are even numbers, d) since infinite symmetric pairs of quanta can be placed in the same cartesian axes and infinite numbers of them can be symmetrized, these pairs have to be bosons.

\section{The U(6) symmetry}\label{space}

The Shell Model $U(3)$ symmetry contains symmetric pairs of quanta in the same cartesian axis, {\it i.e.}, 
\begin{gather}
a_x^\dagger(i_w) a_x^\dagger(i_{w'}),\qquad a_y^\dagger(i_w) a_y^\dagger(i_{w'}),\qquad  a_z^\dagger(i_w) a_z^\dagger(i_{w'}).\label{3}
\end{gather} 
It is not possible to antisymmetrize two quanta in the same cartesian axis, because they cancel out.
The above pairs of quanta are represented by the Young tableaux:
\begin{equation}\label{12new}
\begin{Young}
{\bf x}&{\bf x}\cr
\end{Young}\qquad
\begin{Young}
{\bf y}&{\bf y}\cr
\end{Young}\qquad
\begin{Young}
{\bf z}&{\bf z}\cr
\end{Young}
\end{equation}
respectively. 

But pairs of quanta in different cartesian axes are also possible, which result from the actions:
\begin{gather}
a_x^\dagger(i_w) a_x^\dagger(i_{w'}),\qquad a_x^\dagger(i_w) a_y^\dagger(i_{w'}),\qquad a_x^\dagger(i_w) a_z^\dagger(i_{w'}),\nonumber\\
a_y^\dagger(i_w) a_x^\dagger(i_{w'}),\qquad a_y^\dagger(i_w) a_y^\dagger(i_{w'}),\qquad a_y^\dagger(i_w) a_z^\dagger(i_{w'}), \nonumber\\
a_z^\dagger(i_w) a_x^\dagger(i_{w'}),\qquad a_z^\dagger(i_w) a_y^\dagger(i_{w'}),\qquad a_z^\dagger(i_w) a_z^\dagger(i_{w'}),\nonumber\\\label{9}
\end{gather} 
on the vacuum $|0)$.

The algebraic structure, which follows from all these possible couplings is the:
\begin{gather}
U(3)\otimes U(3)=U(9)\supset U(6).
\end{gather}
In the above $U(3)$ stands for the Shell Model $U(3)$ algebra, where 3 signifies either the catresian quanta in the $z,x,y$ axes, or the spherical quanta in the $\mathcal{m}=\pm 1, 0$ projections of the orbital angular momentum $l=1$. The product $U(3)\otimes U(3)=U(9)$ stands for the coupling of two cartesian quanta into every possible symmetric and antisymmetric pair.

But since the pairs of the cartesian quanta are symmetric upon their interchange, there are some equivalent operators:
\begin{gather}
a_x^\dagger(i_w) a_y^\dagger(i_{w'})=a_y^\dagger(i_w) a_x^\dagger(i_{w'}),\nonumber\\
a_x^\dagger(i_w) a_z^\dagger(i_{w'})=a_z^\dagger(i_w) a_x^\dagger(i_{w'}),\nonumber\\
a_y^\dagger(i_w) a_z^\dagger(i_{w'})=a_z^\dagger(i_w) a_y^\dagger(i_{w'}).\label{3a}
\end{gather}
Consequently out of the 9 pairs formed by the operators (\ref{9}) only the six of them remain, when the discussion is about the symmetric pairs of quanta:
\begin{gather}
\mbox{6 dimensional space}\nonumber\\
a_x^\dagger(i_w) a_x^\dagger(i_{w'}),\qquad a_x^\dagger(i_w) a_y^\dagger(i_{w'}),\qquad a_x^\dagger(i_w) a_z^\dagger(i_{w'}),\nonumber\\
a_y^\dagger(i_w) a_y^\dagger(i_{w'}),\qquad a_y^\dagger(i_w) a_z^\dagger(i_{w'}),\qquad
a_z^\dagger(i_w) a_z^\dagger(i_{w'}).\label{6}
\end{gather} 
The conjugate of them are:
\begin{gather}
a_x(i_{w'}) a_x(i_w),\qquad a_y(i_{w'}) a_x(i_w),\qquad a_z(i_{w'}) a_x(i_w),\nonumber\\
a_y(i_{w'}) a_y(i_w),\qquad a_z(i_{w'}) a_y(i_w),\qquad
a_z(i_{w'}) a_z(i_w).\label{6con}
\end{gather} 
The Young tableaux of the symmetric pairs of quanta in the 6 dimensional space shall be the:
\begin{gather}\label{6new}
\begin{Young}
{\bf x}&{\bf x}\cr
\end{Young}=
\begin{Young}
{\bf xx}\cr
\end{Young}
\mbox{ },\qquad
\begin{Young}
{\bf x}&{\bf y}\cr
\end{Young}=
\begin{Young}
{\bf xy}\cr
\end{Young}\mbox{ },\qquad 
\begin{Young}
{\bf x}&{\bf z}\cr
\end{Young}=
\begin{Young}
{\bf xz}\cr
\end{Young}\mbox{ },
\nonumber\\
\begin{Young}
{\bf y}&{\bf y}\cr
\end{Young}=
\begin{Young}
{\bf yy}\cr
\end{Young}\mbox{ },\qquad
\begin{Young}
{\bf y}&{\bf z}\cr
\end{Young}=
\begin{Young}
{\bf yz}\cr
\end{Young}\mbox{ },\qquad
\begin{Young}
{\bf z}&{\bf z}\cr
\end{Young}=
\begin{Young}
{\bf zz}\cr
\end{Young}\mbox{ }.
\end{gather}

In Eq. (\ref{Fock}) we had defined the states $|n_{kk'})$, which contain $n_{kk'}$ number of symmetric pairs of quanta in the $k,k'$ cartesian axes. It is time to define the Fock state:
\begin{gather}
|n_{xx},n_{xy},n_{xz},n_{yy},n_{yz},n_{zz})
\end{gather}
of the 6 dimensional space with obvious meaning.  If $(k,k')\ne (k'',k''')$ the action:
\begin{gather}
\left(a_k^\dagger(i_w) a_{k'}^\dagger(i_{w'})\right) \left(a_{k''}^\dagger(i_{w''}) a_{k'''}^\dagger(i_{w'''})\right)|n_{kk'},n_{k''k'''}),\label{kk'''}
\end{gather}
with $w=2n_{kk'}+1, w'=2n_{kk'}+2$, $w''=2n_{k''k'''}+1,w'''=2n_{k''k'''}+2$, is about the creation of a symmetric pair of quanta (the $w,w'$) in the $k,k'$ cartesian axes and a symmetric pair of quanta (the $w'',w'''$) in the $k'',k'''$ axes, where the $(k,k'),(k'',k''')$ get the values $(x,x),(x,y),(x,z),(y,y),(y,z),(z,z)$. The relevant Young tableau shall be the:
\begin{gather}
\begin{Young}
${\bf k}$ & ${\bf k'}$\cr
$\bf k''$ & ${\bf k'''}$\cr
\end{Young}\label{4new}
\end{gather}
which means that the quanta in the $k,k'$ axes are symmetric upon their interchange, the same stands for the quanta in the $k'',k'''$ axes, but the pair in the $(k,k')$ axes is antisymmetric with the pair of the $(k'',k''')$ axes. The state (\ref{kk'''}) is equal to the:
\begin{gather}
\left(a_{k''}^\dagger(i_{w''}) a_{k'''}^\dagger(i_{w'''})\right) \left(a_{k}^\dagger(i_{w}) a_{k'}^\dagger(i_{w'})\right)|n_{kk'},n_{k''k'''}),\label{k'''k}
\end{gather}
which in other words means that:
\begin{gather}
[a_k^\dagger(i_w) a_{k'}^\dagger(i_{w'}), a_{k''}^\dagger(i_{w''}) a_{k'''}^\dagger(i_{w'''})]=0.\label{com4}
\end{gather}
The conjugate of the above is:
\begin{gather}
[a_{k'}(i_{w'}) a_{k}(i_{w}), a_{k'''}(i_{w'''}) a_{k''}(i_{w''})]=0.\label{com5}
\end{gather}

One more commutator is left to compute:
\begin{gather}
[a_{k'}(i_{w'})a_{k}(i_{w}),a_{k''}^\dagger(i_{w''})a_{k'''}^\dagger(i_{w'''})]|n_{kk'},n_{k''k'''}),
\end{gather}
with $w'=2n_{kk'},w=2n_{kk'}-1,w''=2n_{k''k'''}+1,w'''=2n_{k''k'''}+2$.
If we utilize the actions of Eqs. (\ref{act1}), (\ref{act2}) we get:
\begin{gather}
[a_{k'}(i_{w'})a_{k}(i_{w}),a_{k''}^\dagger(i_{w''})a_{k'''}^\dagger(i_{w'''})]|n_{kk'},n_{k''k'''})=\nonumber\\
\delta_{k,k''}\delta_{k',k'''}|n_{kk'},n_{k''k'''})
\end{gather}
or:
\begin{gather}
[a_{k'}(i_{w'})a_{k}(i_{w}),a_{k''}^\dagger(i_{w''})a_{k'''}^\dagger(i_{w'''})]=
\delta_{k,k''}\delta_{k',k'''}.\label{com6}
\end{gather}

The commutators (\ref{com4}), (\ref{com5}), (\ref{com6}) ensure that this 6 dimensional space of the symmetric pairs of quanta is a boson space. If $n_{xx}\ge n_{xy}\ge n_{xz}\ge n_{yy}\ge n_{yz}\ge n_{zz}$, the the Young tableau in this 6 dimensional space looks like the:
\begin{gather}
\begin{Young}
$\bf xx$& ${\bf xx}$ & {\bf xx}& {\bf xx}&${\bf xx}$&${\bf xx}$&${\bf xx}$& ... & {\bf xx}\cr
$\bf xy$& $\bf xy$& {\bf xy}& {\bf xy}&${\bf xy}$&${\bf xy}$& ... & {\bf xy}\cr
$\bf xz$& $\bf xz$&{\bf xz}& {\bf xz}&$\bf xz$& ... & {\bf xz}\cr
$\bf yy$&{\bf yy}& {\bf yy}& ... & {\bf yy}\cr
$\bf yz$& {\bf yz}& ... & {\bf yz}\cr
{\bf zz}& ... & {\bf zz}\cr
\end{Young}\label{6new}
\end{gather}
The number of boxes in each line is equal to $f_1,f_2,f_3$, $f_4,f_5,f_6$ respectively. These numbers represent the irrep $[f_1,f_2,f_3,f_4,f_5,f_6]$ of the specific $U(6)$ symmetry. While in the ordinary Shell Model $U(3)$ symmetry it is meaningless to antisymmetrize two quanta in the same cartesian axis (because they cancel out), now in the $U(6)$ symmetry of the symmetric pairs of quanta we may antisymmetrize a pair in the $(k,k')$ axes with a pair in the $(k,k'')$ axes. For instance an antisymmetric pair:
\begin{gather}
\begin{Young}
${\bf xx}$\cr
$\bf xy$\cr
\end{Young}\label{2new}
\end{gather}
exists. The interested reader might try it as an exercise.

The generators of this $U(6)$ algebra have the form:
\begin{gather}
\left(a_k^\dagger(i_w) a_{k'}^\dagger(i_{w'}) \right)\left(a_{k''}(i_w)a_{k'''}(i_{w'}) \right),\label{gen6}
\end{gather}
where the $k,k',k'',k'''$ obtain the values $x,y,z$. There are 36 generators of the above type and $36 \times 36=1296$ commutators among them. The commutators can be computed using the identity (\ref{id6}) and the commutation relations (\ref{com4}), (\ref{com5}), (\ref{com6}) for $w''=w,w'''=w'$. The operators (\ref{gen6}) close upon commutation the $U(6)$ algebra.

\section{The Interacting Boson Model}\label{IBM}

 The interesting thing which has occurred is that {\it anything}, which is constructed by the cartesian operators $a_k(i_w),a_k^\dagger(i_w)$ in the Shell Model $SU(3)$ symmetry, can be equally constructed by the spherical operators $u_\mathcal{m}(i_w),u_\mathcal{m}^\dagger(i_w)$ of Eqs. (\ref{pd-})-(\ref{pd+}).

Since the $u_\mathcal{m}^\dagger(i_w)$ are spherical tensors of degree 1, (see the Appendix B), one may couple a pair of them (see Eq. (\ref{couple})), to create a spherical tensor of degree:\\
a) $\mathcal{L}=0$,\\
b) $\mathcal{L}=1$,\\
c) $\mathcal{L}=2$.\\
According to Eq. (\ref{couple}) we may define the spherical operator ${F^\mathcal{L}_\mathcal{M}}^\dagger(i_w,i_{w'})$, which creates a symmetric pair of quanta with angular momentum $\mathcal{L}$ and projection $\mathcal{M}$ deriving from the $i_w,i_{w'}$ particles:
\begin{gather}
{F^\mathcal{L}_\mathcal{M}}^\dagger(i_w,i_{w'})=\sum_{\mathcal{m},\mathcal{m'}}(\mathcal{1m1m'}|\mathcal{LM})u_\mathcal{m}^\dagger(i_w)  u_\mathcal{m'}^\dagger(i_{w'}).\label{Fd}
\end{gather}
The symbol $F^\mathcal{L}_\mathcal{M}$ is inspired from the Interacting Two Vector Boson Model \cite{Raychev1972,Raychev1978,Georgieva1982}.

Usually a spherical tensor of degree $0, 1, 2$ is called $\mathcal{s^\dagger,p^\dagger,d^\dagger}$ operator respectively. For the $\mathcal{s^\dagger,d^\dagger}$ we define:
\begin{gather}
\mathcal{s}^\dagger={F^0_0}^\dagger,\qquad
\tilde{\mathcal{s}}=(\mathcal{s}^\dagger)^\dagger,\label{sdef}\\
\mathcal{d_M}^\dagger={F^2_\mathcal{M}}^\dagger,\qquad \tilde{\mathcal{d_M}}=(-1)^{\mathcal{M}}(\mathcal{d}_{-\mathcal{M}}^\dagger)^\dagger.\label{ddef}
\end{gather}
The $\mathcal{s}^\dagger$, $\mathcal{d_M}^\dagger$ operators create a symmetrized pair of spherical harmonic oscillator quanta upon permutation with angular momentum 0 and 2 respectively, when acting on the vacuum $|0)$. The tilde operators, in order to ensure that the $\tilde{F}_{\mathcal{M}}^\mathcal{L}$ are spherical tensors,  follow the relation:
\begin{equation}\label{tilde}
\tilde{F}_{\mathcal{M}}^\mathcal{L}=(-1)^{\mathcal{L-M}}\left({F_\mathcal{-M}^\mathcal{L}}^\dagger\right)^\dagger.
\end{equation}
Accordingly a $\mathcal{p}^\dagger$ operator could be defined as:
\begin{equation}\label{pop}
\mathcal{p_M}^\dagger={F^1_\mathcal{M}}^\dagger.
\end{equation}
This $\mathcal{p}^\dagger|0)$  would be a symmetrized pair of quanta with angular momentum 1, but the action of the $\mathcal{p_M}^\dagger$ on the vacuum gives zero. Thus it is meaningless to define such an operator for symmetric pairs of quanta and only the definition of the $\mathcal{s,d}$ operators is worthy, when we need symmetric pairs of quanta. 

On the contrary it would be meaningful to define an antisymmetric pair of quanta with $\mathcal{L}=1$. Such a pair would have a negative parity and would help to predict negative parity ($K^-$) bands in even-even nuclei. But the negative parity bands are not in the purposes of the present article, thus I shall leave the definition of the 3 types ($\mathcal{p}_{-1}^\dagger$, $\mathcal{p}_0^\dagger$, $\mathcal{p}_1^\dagger$) of the antisymmetric pairs of quanta for the future. 

Explicitly, through the calculation of the Clebsch-Gordan coefficients of Eq. (\ref{Fd}), the new creation operators for a pair of spherical quanta result to the expressions:
\begin{gather}
\mathcal{s}^\dagger(i_w,i_{w'})=\nonumber\\
{1\over \sqrt{3}}\left(u_1^\dagger(i_w) u_{-1}^\dagger(i_{w'})+u_{-1}^\dagger(i_w) u_1^\dagger(i_{w'})-u_0^\dagger(i_w) u_0^\dagger(i_{w'})\right),\label{sd}
\end{gather}
\begin{gather}
\mathcal{d}_{-2}^\dagger(i_w,i_{w'})=u_{-1}^\dagger(i_{w}) u_{-1}^\dagger(i_{w'}),\label{dd-2}
\end{gather}
\begin{gather}
\mathcal{d}_{-1}^\dagger(i_w,i_{w'})=\nonumber\\
{1\over\sqrt{2}}\left(u_{-1}^\dagger(i_w) u_0^\dagger(i_{w'})+u_0^\dagger(i_w) u_{-1}^\dagger(i_{w'})\right)\label{dd-1}
\end{gather}
\begin{gather}
\mathcal{d}_0^\dagger(i_w,i_{w'})=
{1\over\sqrt{6}}\left(u_1^\dagger(i_w) u_{-1}^\dagger(i_{w'})+u_{-1}^\dagger(i_w) u_1^\dagger(i_{w'})\right)\nonumber\\
+\sqrt{2\over 3}u_0^\dagger(i_w) u_0^\dagger(i_{w'}),\label{dd0}
\end{gather}
\begin{gather}
\mathcal{d}_1^\dagger(i_w,i_{w'})={1\over\sqrt{2}}\left(u_1^\dagger(i_w) u_0^\dagger(i_{w'})+u_0^\dagger(i_w) u_1^\dagger(i_{w'})\right),\label{dd1}
\end{gather}
\begin{gather}
\mathcal{d}_2^\dagger(i_w,i_{w'})=u_1^\dagger(i_w) u_1^\dagger(i_{w'}),\label{dd2}
\end{gather}
while, following the identity $(AB)^\dagger=B^\dagger A^\dagger$ and Eq. (\ref{tilde}), we get the tilde annihilation operators. 

The same operators can be written in terms of the cartesian operators using the correspondence of the Eqs. (\ref{pd-})-(\ref{pd+}):
\begin{gather}
\mathcal{s}^\dagger(i_w,i_{w'})=\nonumber\\
-{1\over \sqrt{3}}\Big(a_x^\dagger(i_w) a_x^\dagger(i_{w'})+a_y^\dagger(i_w) a_y^\dagger(i_{w'})+a_z^\dagger(i_w) a_z^\dagger(i_{w'})\Big),\label{sdcart}
\end{gather}
\begin{gather}
\mathcal{d}_{-2}^\dagger(i_w,i_{w'})=
{1\over 2}\Big(a_x^\dagger(i_w) a_x^\dagger(i_{w'})-a_y^\dagger(i_w) a_y^\dagger(i_{w'})\nonumber\\
-i\big(a_x^\dagger(i_w) a_y^\dagger(i_{w'})+a_y^\dagger(i_w) a_x^\dagger(i_{w'})\big)\Big),\label{d-2dcart}
\end{gather}
\begin{gather}
\mathcal{d}_{-1}^\dagger(i_w,i_{w'})={1\over 2}\Big(a_x^\dagger(i_w) a_z^\dagger(i_{w'}) +a_z^\dagger(i_w) a_x^\dagger(i_{w'})\nonumber\\
-i\big(a_y^\dagger(i_w) a_z^\dagger(i_{w'})+a_z^\dagger(i_w) a_y^\dagger(i_{w'})\big)\Big),\label{d-1cart}
\end{gather}
\begin{gather}
\mathcal{d}_0^\dagger(i_w,i_{w'})=-{1\over \sqrt{6}}\Big(a_x^\dagger(i_w) a_x^\dagger(i_{w'})+a_y^\dagger(i_w) a_y^\dagger(i_{w'})\Big)\nonumber\\
+\sqrt{2\over 3}a_z^\dagger(i_w) a_z^\dagger(i_{w'}),\label{d0cart}
\end{gather}
\begin{gather}
\mathcal{d}_1^\dagger(i_{w},i_{w'})=-{1\over 2}\Big(a_x^\dagger(i_w) a_z^\dagger(i_{w'})+a_z^\dagger(i_w) a_x^\dagger(i_{w'})\nonumber\\
+i\big(a_y^\dagger(i_w) a_z^\dagger(i_{w'})+a_z^\dagger(i_{w}) a_y^\dagger(i_{w'})\big)\Big),
\end{gather}
\begin{gather}
\mathcal{d}_2^\dagger(i_w,i_{w'})={1\over 2}\Big(a_x^\dagger(i_w) a_x^\dagger(i_{w'})-a_y^\dagger(i_w) a_y^\dagger(i_{w'})\nonumber\\
+i\big(a_x^\dagger(i_w) a_y^\dagger(i_{w'})+a_y^\dagger(i_w) a_x^\dagger(i_{w'})\big)\Big).\label{d2dcart}
\end{gather}

The above operators had been constructed using the pairs of quanta of Eq. (\ref{9}). We may construct the same operators in the 6 dimensional space of Eq. (\ref{6}) by using the relations (\ref{3a}). For instance the $\mathcal{d}_{-2}^\dagger$ operator is given by Eq. (\ref{d-2dcart}), but since: 
\begin{gather}
a_x^\dagger(i_w) a_y^\dagger(i_{w'})|0)=a_y^\dagger(i_w) a_x^\dagger(i_{w'})|0)=\nonumber\\
{1\over \sqrt{2}}\Big(\ket{1_x(i_w),1_y(i_{w'})}+\ket{1_x(i_{w'}),1_y(i_w)}\Big),
\end{gather}
the action:
\begin{gather}
\Big(a_x^\dagger(i_w) a_y^\dagger(i_{w'})+a_y^\dagger(i_w) a_x^\dagger(i_{w'})\Big)|0)=\nonumber\\
={1\over \sqrt{2}}\Big(\ket{1_x(i_w),1_y(i_{w'})}+\ket{1_x(i_{w'}),1_y(i_w)}\Big)\nonumber\\
+{1\over \sqrt{2}}\Big(\ket{1_y(i_w),1_x(i_{w'})}+\ket{1_y(i_{w'}),1_x(i_w)}\Big)
\end{gather}
yields a probability $1$ for the state $\ket{1_x(i_w),1_y(i_{w'})}$ and $1$ for the $\ket{1_x(i_{w'}),1_y(i_w)}$. These probabilities are conserved, if we substitute:
\begin{gather}
\Big(a_x^\dagger(i_w) a_y^\dagger(i_{w'})+a_y^\dagger(i_w) a_x^\dagger(i_{w'})\Big)|0)=\nonumber\\
\sqrt{2}a_x^\dagger(i_w) a_y^\dagger(i_{w'})|0)=\nonumber\\
\ket{1_x(i_w),1_y(i_{w'})}+\ket{1_x(i_{w'}),1_y(i_w)}.
\end{gather}
Similarly:
\begin{gather}
\Big(a_x^\dagger(i_w) a_z^\dagger(i_{w'})+a_z^\dagger(i_w) a_x^\dagger(i_{w'})\Big)|0)=\nonumber\\
\sqrt{2}\big(a_x^\dagger(i_w) a_z^\dagger(i_{w'})\big)|0),
\end{gather}
\begin{gather}
\Big(a_y^\dagger(i_w) a_z^\dagger(i_{w'})+a_z^\dagger(i_w) a_y^\dagger(i_{w'})\Big)|0)=\nonumber\\
\sqrt{2}\big(a_y^\dagger(i_w) a_z^\dagger(i_{w'})\big)|0).
\end{gather}

The 6 dimensional space, we introduced in (\ref{6}), was expressed in terms of the symmetric pairs of the cartesian harmonic oscillator quanta. The same space can be expressed in terms of the 6 spherical tensor operators $\mathcal{s}^\dagger, \mathcal{d_M}^\dagger$. In this sense we may write down the operators $\mathcal{s^\dagger,d^\dagger}$ once more, but now in the 6 dimensional space:
\begin{gather}
\mathcal{s}^\dagger(i_w,i_{w'})=\nonumber\\
-{1\over \sqrt{3}}\Big(a_x^\dagger(i_w) a_x^\dagger(i_{w'})+a_y^\dagger(i_w) a_y^\dagger(i_{w'})+a_z^\dagger(i_w) a_z^\dagger(i_{w'})\Big),\label{sdcart6}
\end{gather}
\begin{gather}
\mathcal{d}_{-2}^\dagger=\nonumber\\
{1\over 2}\Big(a_x^\dagger(i_w) a_x^\dagger(i_{w'})-a_y^\dagger(i_w) a_y^\dagger(i_{w'})-{i\sqrt{2}}a_x^\dagger(i_w) a_y^\dagger(i_{w'})\Big),\label{d-2dcart6}
\end{gather}
\begin{gather}
\mathcal{d}_{-1}^\dagger={1\over\sqrt{2}}\Big(a_x^\dagger(i_w) a_z^\dagger(i_{w'}) -{i}a_y^\dagger(i_w) a_z^\dagger(i_{w'})\Big),
\end{gather}
\begin{gather}
\mathcal{d}_0^\dagger=-{1\over \sqrt{6}}\Big(a_x^\dagger(i_w) a_x^\dagger(i_{w'})+a_y^\dagger(i_w) a_y^\dagger(i_{w'})\Big)
\nonumber\\
+\sqrt{2\over 3}a_z^\dagger(i_w) a_z^\dagger(i_{w'}),
\end{gather}
\begin{gather}
\mathcal{d}_1^\dagger=-{ 1\over\sqrt{2}}\Big(a_x^\dagger(i_w) a_z^\dagger(i_w)+ia_y^\dagger(i_w) a_z^\dagger(i_{w'})\Big),
\end{gather}
\begin{gather}
\mathcal{d}_2^\dagger=\nonumber\\
{1\over 2}\Big(a_x^\dagger(i_w) a_x^\dagger(i_{w'})-a_y^\dagger(i_w) a_y^\dagger(i_{w'})+{i \sqrt{2}}a_x^\dagger(i_w) a_y^\dagger(i_{w'})\Big).\label{d2dcart6}
\end{gather}

Accordingly their conjugates in the 6 dimensional space are:
\begin{gather}
\mathcal{s}=\nonumber\\
-{1\over \sqrt{3}}\Big(a_x(i_{w'}) a_x(i_w)+a_y(i_{w'}) a_y(i_w)+a_z(i_{w'})a_z(i_w)\Big),\label{scart6}
\end{gather}
\begin{gather}
\mathcal{d}_{-2}=\nonumber\\
{1\over 2}\Big(a_x(i_{w'}) a_x(i_w)-a_y(i_{w'}) a_y(i_w)+{i\sqrt{2}}a_x(i_{w'}) a_y(i_w)\Big),\label{d-2cart6}
\end{gather}
\begin{gather}
\mathcal{d}_{-1}={1\over\sqrt{2}}\Big(a_x(i_{w'})a_z(i_w) +{i}a_y(i_{w'}) a_z(i_w)\Big),
\end{gather}
\begin{gather}
\mathcal{d}_0=-{1\over \sqrt{6}}\Big(a_x(i_{w'}) a_x(i_w)+a_y(i_{w'}) a_y(i_w)\Big)
\nonumber\\
+\sqrt{2\over 3}a_z(i_{w'}) a_z(i_w),
\end{gather}
\begin{gather}
\mathcal{d}_1=-{ 1\over\sqrt{2}}\Big(a_x(i_{w'}) a_z(i_w)-ia_y(i_{w'}) a_z(i_w)\Big),
\end{gather}
\begin{gather}
\mathcal{d}_2=\nonumber\\
{1\over 2}\Big(a_x(i_{w'}) a_x(i_w)-a_y(i_{w'})a_y(i_w)-{i \sqrt{2}}a_x(i_{w'}) a_y(i_w)\Big),\label{d2cart6}
\end{gather}
where we had used the symmetry relations (\ref{3a}) and the $(AB)^\dagger=B^\dagger A^\dagger$.

The above relations reveal that there is a unitary transformation in the 6 dimensional space, which transforms the symmetric, cartesian pairs of quanta to the symmetric, spherical pairs of quanta possessing angular momentum $\mathcal{L}=0, 2$. This unitary transformation among the cartesian and spherical symmetric pairs of quanta is the $L$-projection technique, which was introduced by Elliott and Harvey in Ref. \cite{Elliott3} and was accomplished later on in Ref. \cite{Vergados1968}.

We may calculate the commutators among the spherical tensor operators with the use of the Eqs. (\ref{sdcart6})-(\ref{d2cart6}), (\ref{com4}), (\ref{com5}), (\ref{com6}) and the identities (\ref{id1}), (\ref{id3}). The resulting commutators are the:
\begin{gather}
[\mathcal{s}(i_w,i_{w'}),\mathcal{s}^\dagger(i_w,i_{w'})]=1,\\
[\mathcal{d_M}(i_w,i_{w'}),\mathcal{d_{M'}}^\dagger(i_w,i_{w'})]=\delta_{\mathcal{M,M'}},\\
[\mathcal{s}(i_w,i_{w'}),\mathcal{d_M}^\dagger(i_w,i_{w'})]=0,\label{ssd}\\
[\mathcal{s}(i_w,i_{w'}),\mathcal{s}(i_{w''},i_{w'''})]=0\\
[\mathcal{d_M}(i_w,i_{w'}),\mathcal{d_{M'}}(i_{w''},i_{w'''})]=0
\\
[\mathcal{s}(i_w,i_{w'}),\mathcal{d_M}(i_{w''},i_{w'''})]=0,\label{ss}
\\
[\mathcal{s}^\dagger(i_w,i_{w'}),\mathcal{s}^\dagger(i_{w''},i_{w'''})]=0
\\
[\mathcal{d_M}^\dagger(i_w,i_{w'}),\mathcal{d_{M'}}^\dagger(i_{w''},i_{w'''})]=0\\
[\mathcal{s}^\dagger(i_w,i_{w'}),\mathcal{d_M}^\dagger(i_{w''},i_{w'''})]=0.\label{sdsd}
\end{gather}

Therefore the $\mathcal{s}^\dagger, \mathcal{d_M}^\dagger$ operators when acting on the vacuum $|0)$ create a boson, which is a symmetric pair of harmonic oscillator quanta possessing angular momentum $\mathcal{L}=0,2$ respectively. Similarly to Eq. (\ref{Fock}) a state which consists by $n_{F^\mathcal{L}_\mathcal{M}}$ pairs of bosons is constructed by the action:
\begin{gather}
|n_{F^\mathcal{L}_\mathcal{M}}=N_{F^\mathcal{L}_\mathcal{M}})=\nonumber\\{1\over\sqrt{N_{F^\mathcal{L}_\mathcal{M}}!}}\Big({F^\mathcal{L}_\mathcal{M}}^\dagger(i_{2n_{F^\mathcal{L}_\mathcal{M}}+1},i_{2n_{F^\mathcal{L}_\mathcal{M}}+2})\Big)^{N_{F^\mathcal{L}_\mathcal{M}}}|n_{F^\mathcal{L}_\mathcal{M}}=0).
\end{gather}
The Fock space, in which the numbers $n_{F^\mathcal{L}_\mathcal{M}}$ of the $s,d$ bosons are counted may be labeled as:
\begin{gather}
|N)=|n_s,n_{d_{-2}},n_{d_{-1}},n_{d_0},n_{d_1},n_{d_2}).
\end{gather}

Similarly to the actions (\ref{act1}) and (\ref{act2}) the actions of the spherical operators are:
\begin{gather}
\mathcal{s}^\dagger(i_w,i_{w'}) |n_{s})=\sqrt{n_{s}+1}|n_{s}+1),\label{actionsd}\\
\mathcal{d_{-2}}^\dagger(i_w,i_{w'}) |n_{d_{-2}})=\sqrt{n_{d_{-2}}+1}|n_{d_{-2}}+1),\\
\mathcal{d_{-1}}^\dagger(i_w,i_{w'}) |n_{d_{-1}})=\sqrt{n_{d_{-1}}+1}|n_{d_{-1}}+1),\\
\mathcal{d_{0}}^\dagger(i_w,i_{w'}) |n_{d_{0}})=\sqrt{n_{d_{0}}+1}|n_{d_{0}}+1),\\
\mathcal{d_{1}}^\dagger(i_w,i_{w'}) |n_{d_{1}})=\sqrt{n_{d_{1}}+1}|n_{d_{1}}+1),\\
\mathcal{d_{2}}^\dagger(i_w,i_{w'})|n_{d_2})=\sqrt{n_{d_2}+1}|n_{d_2}+1),
\end{gather}
where $w=2n_{F^\mathcal{L}_\mathcal{M}}+1$, $w'=2n_{F^\mathcal{L}_\mathcal{M}}+2$. The annihilator operators act according to the:
\begin{gather}
\mathcal{s}(i_{w'},i_{w}) |n_{s})=\sqrt{n_{s}}|n_{s}+1),\\
\mathcal{d_{-2}}(i_{w'},i_{w}) |n_{d_{-2}})=\sqrt{n_{d_{-2}}}|n_{d_{-2}}+1)\\
\mathcal{d_{-1}}(i_{w'},i_{w}) |n_{d_{-1}})=\sqrt{n_{d_{-1}}}|n_{d_{-1}}+1)\\
\mathcal{d_{0}}(i_{w'},i_{w}) |n_{d_{0}})=\sqrt{n_{d_{0}}}|n_{d_{0}}+1)\\
\mathcal{d_{1}}(i_{w'},i_{w}) |n_{d_{1}})=\sqrt{n_{d_{1}}}|n_{d_{1}}+1)\\
\mathcal{d_{2}}(i_{w'},i_{w})|n_{d_2})=\sqrt{n_{d_2}}|n_{d_2}+1),\label{actions}
\end{gather}
with $w'=2n_{F^\mathcal{L}_\mathcal{M}}$ and $w=2n_{F^\mathcal{L}_\mathcal{M}}-1$ for the reason we explained in Eq. (\ref{conjugate}).

Consequently by using symmetric pairs of harmonic oscillator quanta coming from the valence Shell Model space and possessing definite angular momentum $\mathcal{L}=0,2$ we had constructed a 6 dimensional boson space. This boson space matches perfectly with the space of the IBM. Thus, we may interpret the $s,d$ bosons of the IBM in its $SU(3)$ limit as symmetric pairs of harmonic oscillator quanta of the valence shell, with the angular momentum being a good quantum number for these pairs. These harmonic oscillator quanta refer only to the spatial part of the nuclear wave function. 

In the next sections we shall check if all these formulas, we had obtained, are consistent among them and give consistent results with those in the relevant bibliography. Specifically we shall consider the physical meaning of the generators of the $U(3)$ symmetry in the IBM (the number operator, the angular momentum and the quadrupole moment) and we shall compare them with those of the Shell Model $SU(3)$ symmetry. In addition we are about to present the wave functions of the pairs of quanta in the same cartesian axis and we will compare them with the coherent states of Ginocchio and Kirson \cite{Ginocchio1980a,Ginocchio1980}. Furthermore we shall use the $\mathcal{s,d}$ bosons in order to do the $L$-projection from the intrisic cartesian many quanta state to the physical states with good angular momentum \cite{Elliott2,Elliott3} and we will compare our results with those of Vergados \cite{Vergados1968}. 

\section{The number operator of the U(6) algebra}\label{numberU(6)}

It might be that the most important generator of the $U(6)$ algebra of the IBM is the number operator \cite{IBMbook}:
\begin{gather}
N=\mathcal{s}^\dagger(i_{2n_{s}+1},i_{2n_{s}+2}) \mathcal{s}(i_{2n_{s}},i_{2n_{s}-1})\nonumber\\
+\sum_{\mathcal{M}}\mathcal{d_M}^\dagger(i_{2n_{d_M}+1},i_{2n_{d_M}+2}) \mathcal{d_M}(i_{2n_{d_M}},i_{2n_{d_M}-1}),\label{numbersd}
\end{gather} 
since this operator carries a very important physical meaning: it counts the number of the $s,d$ bosons when thinking in terms of spherical operators. If we define that the number operator for a boson with $\mathcal{L=0,2}$ and $\mathcal{M}$ is:
\begin{gather}
N_{F^{\mathcal{L}}_{\mathcal{M}}}={F^{\mathcal{L}}_{\mathcal{M}}}^\dagger(i_{2n_{F^{\mathcal{L}}_{\mathcal{M}}}+1},i_{2n_{F^{\mathcal{L}}_{\mathcal{M}}}+2})F^{\mathcal{L}}_{\mathcal{M}}(i_{2n_{F^{\mathcal{L}}_{\mathcal{M}}}},i_{2n_{F^{\mathcal{L}}_{\mathcal{M}}}-1}),\label{nsb}
\end{gather}
then the total number operator of Eq. (\ref{numbersd}) can be written as:
\begin{gather}
N=\sum_{(\mathcal{L}=0,2),\mathcal{M}}N_{F^{\mathcal{L}}_{\mathcal{M}}}.\label{totalN}
\end{gather} 
The eigenvalue of each number operator reveals, if we use the actions (\ref{actionsd})-(\ref{actions}):
\begin{gather}
N_{F^{\mathcal{L}}_{\mathcal{M}}}|n_{F^{\mathcal{L}}_{\mathcal{M}}})=n_{F^{\mathcal{L}}_{\mathcal{M}}}|n_{F^{\mathcal{L}}_{\mathcal{M}}}).
\end{gather}
Thus the eigenvalue of the number operator using the spherical tensors is:
\begin{equation}
N|N)=(n_{s}+n_{d_{-2}}+n_{d_{-1}}+n_{d_0}+n_{d_1}+n_{d_2})|N).\label{valuenumbersd}
\end{equation} 

It is crucial to see, if the number operator as defined in the IBM, counts also the number of the symmetric pairs of quanta. We can check this by substituting Eqs. (\ref{sdcart6})-(\ref{d2cart6}) into (\ref{numbersd}). By doing so, indeed:
\begin{gather}
N=\Big(a_x^\dagger(i_{2n_{xx}+1}) a_x^\dagger(i_{2n_{xx}+2}) \Big) \Big( a_x(i_{2n_{xx}})  a_x(i_{2n_{xx}-1}) \Big)\nonumber\\
+\Big (a_x^\dagger(i_{2n_{xy}+1})  a_y^\dagger(i_{2n_{xy}+2})  \Big) \Big(a_x (i_{2n_{xy}}) a_y(i_{2n_{xy}-1}) \Big)\nonumber\\
+\Big(a_x^\dagger(i_{2n_{xz}+1})  a_z^\dagger(i_{2n_{xz}+2}) \Big)\Big( a_x(i_{2n_{xz}})  a_z(i_{2n_{xz}-1}) \Big)\nonumber\\
+\Big(a_y^\dagger(i_{2n_{yy}+1})  a_y^\dagger(i_{2n_{yy}+2}) \Big) \Big(a_y(i_{2n_{yy}})  a_y(i_{2n_{yy}-1}) \Big)\nonumber\\
+\Big(a_y^\dagger(i_{2n_{yz}+1})  a_z^\dagger(i_{2n_{yz}+2}) \Big) \Big( a_y(i_{2n_{yz}})  a_z(i_{2n_{yz}-1}) \Big)\nonumber\\
+\Big(a_z^\dagger(i_{2n_{zz}+1})  a_z^\dagger(i_{2n_{zz}+2}) \Big)\Big( a_z(i_{2n_{zz}})  a_z(i_{2n_{zz}-1}) \Big).\label{number}
\end{gather} 
If we act the above operator on the $|N)$ and we implement Eq. (\ref{nkk'}) and the conjugate of Eqs. (\ref{3a}) then the eigenvalue of the number operator shall appear:
\begin{gather}
N|N)=\Big(n_{xx}+n_{xy}+n_{xz}+n_{yy}+n_{yz}+n_{zz}\Big)|N)=\nonumber\\
\Big(n_{s}+n_{d_{-2}}+n_{d_{-1}}+n_{0}+n_{1}+n_{2}\Big)|N).
\end{gather}

The first test of the current idea for the microspopic origin of the IBM in its $SU(3)$ limit was successful, since the number operator in the IBM results to have as eigenvalues the number of the symmetric pairs of quanta, which derive by the occupancy of the cartesian Shell Model space by nucleons.

Consequently {\it the number of bosons in this microscopic justification of the IBM is the number of the symmetric pairs of quanta}. In an Shell Model $SU(3)$ wave function given by a $(\lambda,\mu)$ irrep the total number of quanta is $\lambda+2\mu$, among which the $\lambda+\mu$ quanta are symmetric upon their interchange, the $\mu$ are symmetric among them, but the $\mu$ quanta are nor symmetric neither antisymmetric with the rest $\lambda+\mu$ quanta \cite{proxy5}. Thus if the $\lambda+\mu, \mu$ are even numbers, then the number of the symmetric pairs of quanta is:
\begin{equation}
N={\lambda+2\mu\over 2},\mbox{ for }\lambda+\mu, \mu\mbox{ even}.\label{nlm}
\end{equation}
Consequently this interpretation of the $s,d$ bosons introduces a new way of counting the number of quanta. 

\section{The angular momentum}\label{generators}

In this section, it is our purpose to clarify, what is the relation of the angular momentum in the $SU(3)$ limit of the IBM with that of the Shell Model $SU(3)$ algebra. Do they carry the same physical meaning? Are they identical or simply similar?

The angular momentum of the $i^{th}$ nucleon is defined as:
\begin{gather}
{\bf l}={\bf r}\times{\bf p}.
\end{gather}
With the use of the creation and annihilation operators of Eq. (\ref{a}), this angular momentum results to the expressions (see Appendix A) \cite{Lipkin}:
\begin{gather}
l_x=yp_z-p_yz=\mathcal{i}(a_ya_z^\dagger-a_za_y^\dagger),\label{lx}\\
l_y=zp_x-p_zx=\mathcal{i}(a_za_x^\dagger-a_xa_z^\dagger),\label{ly}\\
l_z=xp_y-p_xy=\mathcal{i}(a_xa_y^\dagger-a_ya_x^\dagger).\label{lz}
\end{gather}
The eigenvalues of the $l_z$ operator when acting on a single particle Shell Model orbital are labeled $m_l$ in Ref. \cite{proxy4}. Accordingly the $l^2$ operator has eigenvalues $l(l+1)$. One may find the expectation values of the $l_z,l^2$ operators when acting on the cartesian Shell Model basis by using the transformations of section 2 of Ref. \cite{proxy4}.

It is really interesting that the single nucleon angular momentum can be found by the symmetric coupling of the harmonic oscillator quanta of the orbital, which hosts this nucleon. I shall give two examples for that.

Suppose for instance that the nucleon occupies the spatial cartesian orbital $\ket{n_z,n_x,n_y}$ = $\ket{2,0,0}$. This orbital equals to a superposition of the spherical orbitals
$\ket{n,l,m_l}$, which had been defined in Eq. (3) of Ref. \cite{proxy4}. In one runs the code of Ref. \cite{proxy4}, or calculates on a piece of paper the result of the transformation of Eq. (5) of Ref. \cite{proxy4} s/he will get that:
\begin{gather}
\ket{n_z=2,n_x=0,n_y=0}=\nonumber\\
-{1\over \sqrt{3}}\ket{n=1,l=0,m_l=0}+\sqrt{2\over 3}\ket{n=0,l=2,m_l=0}.\label{above}
\end{gather}
Exactly the same result can be obtained by the symmetric coupling of two harmonic oscillator quanta in the $z$ axis. The procedure is illustrated later on in this article and the comparison of Eqs. (\ref{above}), (\ref{phizz}) reveals that the two equations are identical.

A second example shall be given for the spatial cartesian state $\ket{n_z,n_x,n_y}=$ $\ket{1,1,0}$. If one runs the code of Ref. \cite{proxy4} for the transformation of Eqs. (4) and (5) of Ref. \cite{proxy4} s/he will get:
\begin{gather}
\ket{n_z=1,n_x=1,n_y=0}\nonumber\\
{1\over\sqrt{2}}\Big(\ket{n=0,l=2,m_l=-1}-\ket{n=0,l=2,m_l=1}\Big). \label{zxstate}
\end{gather}
Exactly the same result has been obtained by the symmetric coupling of one quantum in the $z$ axis with one quantum in the $x$ axis. The comparison of Eqs. (\ref{zxstate}), (\ref{phizx}) reveals that indeed they are identical.

Consequently instead of the single nucleon angular momentum, one may use the angular momentum, which results from the symmetric coupling of the harmonic oscillator quanta. This property reveals once more the physical meaning of the $U(\Omega)\supset U(3)$ decomposition, in which the $\Omega$ dimensional space is occupied by nucleons, while the 3 dimensional space is occupied by quanta. The single nucleon angular momentum is decomposed into the angular momenta of the harmonic oscillator quanta. There lies the beauty of Elliott's achievement \cite{Elliott1,Elliott2}: by decomposing the angular momentum of each single nucleon into the angular momenta of the quanta and afterwards by coupling the angular momenta of all the quanta for the many nucleon system, one gets the familiar rotational spectrum, the one we see in the data.

In the Shell Model $SU(3)$ symmetry the nuclear angular momentum operator is the summation (see Eq. (\ref{LS})):
\begin{gather}
{\bf L}=\sum_{i=1}^{A_{val}}{\bf l}(i)\label{manyl}
\end{gather}
for every $i$ nucleon of the proton and neutron valence shells, where $A_{val}$ is the number of the valence nucleons. This angular momentum operator has projection in the $z$ axis the $L_z$ operator and ladder operators the $L_{\pm}$. One set of generators of the Shell Model $SU(3)$ algebra is the:
\begin{equation}
L_0=L_z, L_{\pm}, Q_0, Q_{\pm 1}, Q_{\pm 2}.
\end{equation}
The operators of the angular momentum are parts of a spherical tensor of degree 1, while the $Q_\mathcal{m}$, with $\mathcal{m}=0,\pm1,\pm2$ represent the algebraic quadrupole moment operators and they are the components of a spherical tensor of degree 2 \cite{Elliott2}. 

Since the single nucleon angular momentum can be decomposed into the angular momenta of the quanta of the orbital which hosts this nucleon, the many nucleon angular momentum of Eq. (\ref{manyl}) can be rewritten as:
\begin{gather}
{\bf L}=\sum_{w=1}^{p+q+r}{\bf l}(i_w),
\end{gather}
or in terms of pairs of quanta:
\begin{gather}
{\bf L}=\sum_{w,w'}\big({\bf l}(i_w)+{\bf l}(i_{w'})\big)
\end{gather}
where $w'=w+1$ and the $w$ obtains the values $1,3,5,..,p+q+r-1$. The angular momentum of the pair of quanta is defined:
\begin{gather}
{\bf l}(i_w,i_{w'})={\bf l}(i_w)+{\bf l}(i_{w'}).\label{lzpair}
\end{gather}

The eigenstates of the $l_z(i_w,i_{w'})$ operator when acting on a pair of quanta with $n_{F^\mathcal{L}_\mathcal{M}}\ge 1$ are:
\begin{gather}
l_z(i_{2n_{s}-1},i_{2n_{s}})|n_s)=
0|n_s), \\ 
l_z(i_{2n_{d_{-2}}-1},i_{2n_{d_{-2}}})|n_{d_{-2}})=
-2|n_{d_{-2}}),\\
l_z(i_{2n_{d_{-1}}-1},i_{2n_{d_{-1}}})|n_{d_{-1}})=
-1|n_{d_{-1}}),\\
l_z(i_{2n_{d_{0}}-1},i_{2n_{d_{0}}})|n_{d_{0}})=
0|n_{d_{0}}),\\
l_z(i_{2n_{d_{1}}-1},i_{2n_{d_{1}}})|n_{d_{1}})=
1|n_{d_{1}}),\\
l_z(i_{2n_{d_{2}}-1},i_{2n_{d_{2}}})|n_{d_{2}})=
2|n_{d_{2}}).
\end{gather} 
If $n_{F^\mathcal{L}_\mathcal{M}}\ge 1$ the above can be written in a compact form using the definition (\ref{Fd}):
\begin{gather}
l_z(i_{2n_{F^\mathcal{L}_\mathcal{M}}-1},i_{2n_{F^\mathcal{L}_\mathcal{M}}})|n_{F^\mathcal{L}_\mathcal{M}})=\mathcal{M} |n_{F^\mathcal{L}_\mathcal{M}}).\label{eigenLz}
\end{gather}
In accordance the square of the angular momentum of a pair of quanta has eigenstates the:
\begin{gather}
l^2(i_{2n_{F^\mathcal{L}_\mathcal{M}}-1},i_{2n_{F^\mathcal{L}_\mathcal{M}}})|n_{F^\mathcal{L}_\mathcal{M}})=\mathcal{L}(\mathcal{L}+1)|n_{F^\mathcal{L}_\mathcal{M}})
\end{gather}
for $n_{F^\mathcal{L}_\mathcal{M}}\ge 1$.

Now, in the IBM side the angular momentum operators are \cite{IBMbook}:
\begin{gather}
J_m=\sqrt{10}\sum_{\mathcal{M,M'}}(2\mathcal{M}2\mathcal{M'}|1m)\mathcal{d_M}^\dagger\tilde{\mathcal{d}}_{\mathcal{M'}},m=0,\pm1.\label{JzIBM}
\end{gather}
Using a Clebsch-Gordan coefficients calculator \cite{Edmonds} and the Eq. (\ref{tilde}) we get:
\begin{gather}
J_z=J_0=(0)\mathcal{s}^\dagger\mathcal{s}+(-2)\mathcal{d_{-2}}^\dagger\mathcal{d_{-2}}+(-1)\mathcal{d_{-1}}^\dagger\mathcal{d_{-1}}\nonumber\\
+(0)\mathcal{d_{0}}^\dagger\mathcal{d_{0}}+(1)\mathcal{d_{1}}^\dagger\mathcal{d_{1}}+(2)\mathcal{d_{2}}^\dagger\mathcal{d_{2}}.
\end{gather}
If we interpret the $\mathcal{s,d}$ bosons as symmetric pairs of quanta, then the above can be written in a compact form:
\begin{equation}
J_z=\sum_{\mathcal{L,M}}\mathcal{M}N_{F^\mathcal{L}_{\mathcal{M}}},
\end{equation}
where $\mathcal{M}$ is the eigenvalue of the $l_z(i_{2n_{F^\mathcal{L}_\mathcal{M}}-1},i_{2n_{F^\mathcal{L}_\mathcal{M}}})$ (see Eq. (\ref{eigenLz})), while $N_{F^\mathcal{L}_{\mathcal{M}}}$ is the number operator of Eq. (\ref{nsb}). The above is written as:
\begin{equation}
J_z=l_z(i_{2n_{F^\mathcal{L}_\mathcal{M}}-1},i_{2n_{F^\mathcal{L}_\mathcal{M}}})N,
\end{equation}
where Eqs. (\ref{totalN}), (\ref{eigenLz}) have been used.

Now the relation of the $J$ in the $SU(3)$ limit of the IBM with the traditional (the one we know from Quantum Mechanics) $l$ operator is revealed: the $J$ results from the coupling of two spherical tensors: a) the number operator being a spherical tensor of degree 0, with b) the angular momentum operator of the pair of quanta which is a spherical tensor of degree 1:
\begin{gather}
J_m=\sum_{\mathcal{m}=\pm 1,0}(1\mathcal{m}00|1m)l_\mathcal{m}(i_{2n_{F^\mathcal{L}_\mathcal{M}}-1},i_{2n_{F^\mathcal{L}_\mathcal{M}}})N,\\
J^2= l^2(i_{2n_{F^\mathcal{L}_\mathcal{M}}-1},i_{2n_{F^\mathcal{L}_\mathcal{M}}})N.
\end{gather}

Despite that the $J^2$ operator used in the IBM, seems slightly different than the $L^2$ of the Elliott Model, these two operators have exactly the same action on the $|N)$ states. So the $J^2$ of the IBM is indeed the physical angular momentum operator. This can be proven as an exercise by acting the $J_0$ and the $L_z$ on the spatial Shell Model $SU(3)$ states of Eqs. (\ref{phizz}), (\ref{phixx}), (\ref{phiyy}).

The second test of the current idea for the microscopic justification of the IBM proved successful too, since within this interpretation of the $\mathcal{s,d}$ bosons the angular moment of the IBM is related to the physical angular momentum and furthermore these two operators have exactly the same actions on the intrinsic Shell Model $SU(3)$ states.

\section{The coherent states}\label{coh}

The coherent states were introduced by Ginocchio and Kirson in Refs. \cite{Ginocchio1980a,Ginocchio1980}, in order to link the IBM with the Collective Model of Bohr and Mottelson \cite{BohrII}. The authors defined the boson creation operator as \cite{Ginocchio1980a}:
\begin{gather}
Q^\dagger (\beta,\gamma)=\nonumber\\
{1\over \sqrt{1+\beta ^2}}\left(\mathcal{s}^\dagger+\beta\cos{\gamma} \mathcal{d_0}^\dagger+{1\over \sqrt{2}}\beta\sin{\gamma}(\mathcal{d_{2}}^\dagger+\mathcal{d_{-2}}^\dagger)\right),\label{Gin}
\end{gather}
where $\beta$ is the quadrupole deformation variable, while $\gamma$ is an angle, which shows the kind of deformation (prolate, oblate, spherical). Accordingly the coherent state is defined \cite{Ginocchio1980b}:
\begin{equation}\label{coherent}
\ket{N;\beta,\gamma}={1\over\sqrt{N!}}[Q^\dagger(\beta,\gamma)]^N\ket{0}.
\end{equation}

In the $SU(3)$ limit of the IBM the deformation is equal to $\beta=\sqrt{2}$, while the angle $\gamma$ may adopt any value (see chapter 13 of Ref. \cite{Dennis}). The corresponding shape for selected values of the $\gamma$ is presented in the book of Greiner and Maruhn (see Fig. 6.4 of Ref. \cite{Greiner}). Specifically:
\begin{gather*}
\gamma=0^\circ, \mbox{ prolate with }x=y,\\
\gamma=60^\circ, \mbox{ oblate with }x=z,\\
\gamma=120^\circ, \mbox{ prolate with }y=z,\\
\gamma=180^\circ, \mbox{ oblate with }x=y,\\
\gamma=240^\circ, \mbox{ prolate with }x=z,\\
\gamma=300^\circ, \mbox{ oblate with }y=z.
\end{gather*}

Right after, it will be proven that the Shell Model $SU(3)$ states are the coherent states of Ginocchio and Kirson, if the $s,d$ bosons of the Interacting Bosom Model are interpreted as symmetric pairs of harmonic oscillator quanta.

\section{Some examples}\label{examples}

In the following some basic examples are demonstrated. The Young tableau of a Shell Model $U(3)$ irrep consists by three rows. The boxes in each row represent the number of quanta in the three cartesian axes. Every $SU(3)$ irrep with $\lambda+\mu,\mu$ even, no matter how lengthy or complicated will be, can be built from pairs of quanta. In other words the intrinsic Shell Model $SU(3)$ states can be translated into the $\mathcal{s,d}$ bosons of the IBM.

\subsection{Two quanta in the z axis }

The easiest example is that of 1 proton or neutron in the s, d 3D-HO shell among the magic numbers 8-20, or equivalently in the $2s^{1/2}$, $1d^{3/ 2}$, $1f^{7/2}$ proxy shell \cite{proxy4} among the 14-26 magic numbers. This particle occupies the cartesian state $\ket{n_z,n_x,n_y}=$ $\ket{2,0,0}$ (see Eq. (4) of Ref. \cite{proxy5}), thus it possesses a $U(3)$ irrep $[f_1,f_2,f_3]$= $[\sum_i n_{z}(i), \sum_i n_{x}(i), \sum_i n_{y}(i)]$= $[2,0,0]$ (see Eq. (21) of Ref. \cite{proxy5}) and so the Shell Model $SU(3)$ irrep is the $(\lambda,\mu)=(2,0)$ (see Eqs. (27), (28) of Ref. \cite{proxy5}). The many-quanta $SU(3)$ wave function of this irrep is represented by the quantum-number \cite{Lipas} Young tableaux:
\begin{equation}\label{20}
\begin{Young} \bf z & \bf z \cr \end{Young}\qquad \begin{Young} 1 & 2\cr \end{Young}.
\end{equation}
The quanta 1, 2 are symmetric upon their interchange, so the spatial wave function is:
\begin{gather}\label{Psic}
\Phi_{zz}=a_z^\dagger(i_1) a_z^\dagger(i_2)|0).
\end{gather}

With the use of the operators of Eq. (\ref{au}), the above wave function is written as:
\begin{gather}\label{Psi12}
\Phi_{zz}=u_0^\dagger(i_1) u_0^\dagger(i_2)|0).
\end{gather}
The operators $u_0^\dagger(i_1) u_0^\dagger(i_2)$ can be written in terms of the $\mathcal{s}^\dagger(i_1,i_2)$, $\mathcal{d}_0^\dagger(i_1,i_2)$ operators of Eqs. (\ref{sd}), (\ref{dd0}) as:
\begin{equation}
u_0^\dagger(i_1) u_0^\dagger(i_2)=-{1\over \sqrt{3}}\mathcal{s}^\dagger(i_1,i_2)+\sqrt{2\over 3}\mathcal{d}_0^\dagger(i_1,i_2).
\end{equation}
We could obtain the same result, if we coupled the $u_0^\dagger(i_1)$ with the $u_0^\dagger(i_2)$ using Eq. (\ref{couple2}). So Eq. (\ref{Psi12}) is equal to:
\begin{gather}
\Phi_{zz}={1\over \sqrt{3}}\left(-\mathcal{s}^\dagger(i_1,i_2)+\sqrt{2}\mathcal{d}_0^\dagger(i_1,i_2) \right)|0).\label{phizz}
\end{gather}

In the Elliott wave functions the band label $K$ is equal to the projection of the angular momentum $M$.
Practically this means that the catresian wave function of Eq. (\ref{Psic}) {\it projects} into an wave function with good angular momentum $L$, so as a nuclear state with $L=0,K=0$ (the $\mathcal{s}$ boson) is included into the cartesian wave function with probability ${1\over 3}$ and an $L=2, K=0$ nuclear state (the $\mathcal{d_0}$ boson) lies within the cartesian wave function with probability $2\over 3$.

This procedure is called {\it L-projection} and was introduced by J. P. Elliott in 1958 in Ref. \cite{Elliott2}. The projection operator $P$ is further explained in the Appendix of Ref. \cite{Elliott3}, while the matrix elements of $P$ have been calculated in 1968 by J. D. Vergados in Ref. \cite{Vergados1968}. 

The third test for the validity of the whole idea for the microscopic origin of the IBM in the $SU(3)$ limit shall be the comparison of the projection coefficients using a) the above method and b) the method of Vergados \cite{Vergados1968}. If both methods result to the same projection coefficients, then we will gain further confidence that the construction of the spatial $SU(3)$ wave functions with the use of the symmetric pairs of quanta is correct.

The $L$-projection of the spatial many-quanta cartesian wave function using the traditional method of Elliott, Harvey and Vergados \cite{Elliott2,Elliott3,Vergados1968} will be presented briefly. The matrix elements of the projection operator within the same $K$ nuclear band are:
\begin{equation}
A(K,L,K)=\braket{\Phi|P|\Phi}=|a(K,L)|^2.
\end{equation}
The coefficients $a(K,L)$ are given in Table 2A of Ref. \cite{Vergados1968}. The $SU(3)$ irrep of our example is the $(\lambda,\mu)$= $(2,0)$ and so:
\begin{gather}
a(K=0,L=0)={1\over\sqrt{3}},\\
a(K=0,L=2)=\sqrt{2\over 3}.
\end{gather}
So, with the use of the symmetric pairs of quanta (the $\mathcal{s,d}$ bosons) and with the use of the Elliott, Harvey and Vergados method, we got the {\it same} projection coefficients for the $\Phi_{zz}$ state. This successful result indicates that the method we used for the construction of the spatial Shell Model $SU(3)$ state gave consistent results with the relevant bibliography \cite{Vergados1968}. 

To test further the interpretation of the $\mathcal{s,d}$ bosons as symmetric pairs of quanta we shall compare the $\Phi_{zz}$ state with the coherent states of Ginocchio and Kirson \cite{Ginocchio1980a}. The Eq. (\ref{phizz}) can be written as:
\begin{gather}
\Phi_{zz}={1\over \sqrt{1+\beta^2}}
\Big(-\mathcal{s}^\dagger(i_1,i_2)
+\beta\cos{\gamma} \mathcal{d_0}^\dagger(i_1,i_2)\nonumber\\
+{1\over \sqrt{2}}\beta\sin{\gamma}\big(\mathcal{d_{2}}^\dagger(i_1,i_2)+\mathcal{d_{-2}}^\dagger(i_1,i_2)\big)\Big)|0)
\end{gather}
with $\beta=\sqrt{2}$ and $\gamma=0^\circ$. Thus, it looks quite similar with the coherent state of Eq. (\ref{coherent}), with the difference that it adopts a negative sign in the $\mathcal{s}^\dagger$ operator. Nevertheless, in the definition of the $\mathcal{s}^\dagger$ operator in Eq. (\ref{sdcart6}), the overall sign is negative, thus the $-\mathcal{s}^\dagger$ is positive. So indeed the $-\mathcal{s}^\dagger$ in this work coincides with the $\mathcal{s}^\dagger$ of Eq. (\ref{Gin}). Therefore the $\Phi_{zz}$ state represents a coherent state of the $SU(3)$ limit of the IBM (since $\beta=\sqrt{2}$) with prolate shape (since $\gamma=0^\circ$). 

Consequently the method we had used for the construction of the $\Phi_{zz}$ state out of the symmetric pairs of quanta (the $\mathcal{s,d}$ bosons) not only gave the correct projection coefficients in comparison with those of Vergados, but also proved that the intrinsic Elliott state with $(\lambda,\mu)=(2,0)$ is a coherent state of the IBM with the correct value of the deformation variable for the $SU(3)$ limit ($\beta=\sqrt{2}$) and the correct value of the angle $\gamma=0^\circ$. This result serves as a successful fourth test for the current justification of the IBM in its $SU(3)$ limit.

\subsection{Two quanta in the x axis }\label{2xx}

Supposing now, that the $SU(3)$ wave function has two quanta in the $x$ cartesian axis this wave function is:
\begin{equation}
\Phi_{xx}=a_x^\dagger(i_1) a_x^\dagger(i_2)|0),
\end{equation}
which, with the use of Eqs. (\ref{au}), can be written as:
\begin{equation}\label{xx}
\Phi_{xx}={u_{-1}^\dagger(i_1)-u_1^\dagger(i_1)\over\sqrt{2}}\cdot{u_{-1}^\dagger(i_2)-u_1^\dagger(i_2)\over\sqrt{2}}|0).
\end{equation}
This wave function is represented by the quantum-number \cite{Lipas} Young tableaux:
\begin{equation}\label{02}
\begin{Young} \bf x &\bf  x \cr \end{Young}\qquad \begin{Young} 1 & 2\cr \end{Young}.
\end{equation}

From Eq. (\ref{couple2}) one gets that:
\begin{gather}
u_{-1}^\dagger(i_1) u_1^\dagger(i_2)={1\over\sqrt{3}}\mathcal{s}^\dagger(i_1,i_2)-{1\over\sqrt{2}}\mathcal{p}_0^\dagger(i_1,i_2)\nonumber\\
+{1\over\sqrt{6}}\mathcal{d}_0^\dagger(i_1,i_2),\\
u_{1}^\dagger(i_1) u_{-1}^\dagger(i_2)={1\over\sqrt{3}}\mathcal{s}^\dagger(i_1,i_2)+{1\over\sqrt{2}}\mathcal{p}_0^\dagger(i_1,i_2)\nonumber\\
+{1\over\sqrt{6}}\mathcal{d}_0^\dagger(i_1,i_2),
\end{gather}
which implies, that:
\begin{gather}
u_{-1}^\dagger(i_1) u_1^\dagger(i_2)+u_1^\dagger(i_1) u_{-1}^\dagger(i_2)=\nonumber\\
{2\over\sqrt{3}}\mathcal{s}^\dagger(i_1,i_2)+{2\over\sqrt{6}}\mathcal{d}_0^\dagger(i_1,i_2).\label{tpt}
\end{gather}
We could obtain the same result with the use of Eqs. (\ref{sd}), (\ref{dd0}).

Thus with the use of the above and of Eqs. (\ref{dd-2}), (\ref{dd2}) the Eq. (\ref{xx}) becomes:
\begin{gather}
\Phi_{xx}={1\over\sqrt{3}}\Big(-\mathcal{s}^\dagger(i_1,i_2)
-{\sqrt{2}\over 2}\mathcal{d}_0^\dagger(i_1,i_2)\nonumber\\
+{\sqrt{3}\over 2}\big(\mathcal{d}_{-2}^\dagger(i_1,i_2)+\mathcal{d}_2^\dagger(i_1,i_2)\big)\Big)|0).\label{phixx}
\end{gather}
In the Elliott $L$-projected wave functions $\psi(KLM)$ \cite{Elliott2} two opposite $K$ states are equal: $\psi(KLM)=\psi(-KLM)$. Thus the $\mathcal{d}_{-2}^\dagger(i_1,i_2)|0)$, $\mathcal{d}_2^\dagger(i_1,i_2)|0)$ refer to the same nuclear state. All these mean, that the wave function (\ref{phixx}) reflects to two nuclear bands with $(K,L)=(0,0), (0,2)$ and $(K,L)=(2,2)$. In Eq. (\ref{phixx}) the state with $K=0,L=0$ appears with probability $1/3$, the $K=0,L=2$ with probability $1/6$ and the $K=2,L=2$ with probability $1/4+1/4=1/2$. The quanta in the $x,y$ plane are responsible for the $\mu$ quantum number (see Eq. (15) of Ref. \cite{Elliott2}). For two quanta in the $x$ axis we get that $\mu=2$ and that $K=0,2$ (see Eq. (22) of Ref. \cite{Elliott2}). So Eq. (\ref{phixx}) predicts correctly the existence of two bands with $K=0,2$.

Furthermore Eq. (\ref{phixx}) can be written as:
\begin{gather}
\Phi_{xx}={1\over \sqrt{1+\beta^2}}
\Big(-\mathcal{s}^\dagger(i_1,i_2)
+\beta\cos{\gamma} \mathcal{d_0}^\dagger(i_1,i_2)\nonumber\\
+{1\over \sqrt{2}}\beta\sin{\gamma}\big(\mathcal{d_{2}}^\dagger(i_1,i_2)+\mathcal{d_{-2}}^\dagger(i_1,i_2)\big)\Big)|0)
\end{gather}
with $\beta=\sqrt{2}$ and $\gamma=120^\circ$. In comparison with Eq. (\ref{coherent}), the $\Phi_{xx}$ represents a coherent state of the $SU(3)$ limit of the IBM with prolate shape and $N=1$. So it represents a prolate shape with equal lengths in the $z,y$ axes, as expected.

\subsection{Two quanta in the $y$ axis }\label{2yy}

If the $SU(3)$ wave function has two quanta in the $y$ cartesian axis, the wave function is:
\begin{equation}
\Phi_{yy}=a_y^\dagger(i_1) a_y^\dagger(i_2) |0),
\end{equation}
which, with the use of Eqs. (\ref{au}), can be written as:
\begin{gather}
\Phi_{yy}=\mathcal{i}{u_{-1}^\dagger(i_1) +u_1^\dagger(i_2) \over\sqrt{2}}\cdot \mathcal{i}{u_{-1}^\dagger(i_1) +u_1^\dagger(i_2) \over\sqrt{2}}|0).\label{yy}
\end{gather}
This wave function is represented by the quantum-number \cite{Lipas} Young tableaux:
\begin{equation}\label{02}
\begin{Young} \bf y & \bf y \cr \end{Young}\qquad \begin{Young} 1 & 2\cr \end{Young}.
\end{equation}

Using the Eqs. (\ref{dd-2}), (\ref{dd2}) and (\ref{tpt}) the Eq. (\ref{yy}) becomes:
\begin{gather}
\Phi_{yy}={1\over\sqrt{3}}\Big(-\mathcal{s}^\dagger(i_1,i_2)-{\sqrt{2}\over 2}\mathcal{d}_0^\dagger(i_1,i_2)\nonumber\\
-{\sqrt{3}\over 2}\big(\mathcal{d}_{-2}^\dagger(i_1,i_2)+\mathcal{d}_2^\dagger(i_1,i_2)\big)\Big)|0).\label{phiyy}
\end{gather}

Equation (\ref{phiyy}) represents once more a coherent state of the $SU(3)$ limit, as defined in Eq. (\ref{coherent}), with $N=1$, $\beta=\sqrt{2}$, $\gamma=240^\circ$. Therefore it represents a prolate shape with equal lengths in the $z,x$ axes.

\subsection{Four quanta in the z axis}

In the previous sections we explored the states of two quanta in the $z$, $x$, $y$ axes. Any other $SU(3)$ wave function with even number of $\lambda+\mu,\mu$ quanta, may be constructed by the outer product $\otimes$ of simple Young tableaux \cite{Lipas,Harvey,Alex2011,Coleman1964,Troltenier1996,Dytrych2021} .

For instance if one proton or neutron is placed in the proxy 50-82 spin-orbit like shell \cite{proxy4} the resulting highest weight irrep is the $(\lambda,\mu)=(4,0)$ \cite{Assimakis,Martinou2021}, which is represented by the Young tableaux:
\begin{equation}
\begin{Young}
\bf z & \bf z & \bf z & \bf z\cr
\end{Young}\qquad
 \begin{Young} 1 & 2 &3 &4\cr \end{Young}
\end{equation}
The above irrep may result from the symmetric product:
\begin{equation}\label{times}
\begin{Young} \bf z & \bf z \cr \end{Young}\mbox{ } \begin{Young} 1 & 2\cr \end{Young} \otimes \begin{Young} \bf z & \bf z\cr \end{Young}\mbox{ }  \begin{Young} 3 & 4\cr \end{Young}.
\end{equation}

The number of the $\mathcal{s,d}$ bosons is given by Eq (\ref{nlm}) to be $N=2$. The wave functions of the two bosons are given by Eq. (\ref{phizz}). Since the quanta are in the same row of the Young tableau they must be symmetric upon their interchange. So the spatial $SU(3)$ wave function shall be a coherent state with two pairs of quanta:
\begin{gather}
\Phi_{space}=
\Big(-{1\over \sqrt{3}}\mathcal{s}^\dagger(i_{2N+1},i_{2N+2})\nonumber\\+\sqrt{2\over 3}\mathcal{d}_0^\dagger(i_{2N+1},i_{2N+2}) \Big)^2|0),
\end{gather}
where Eq. (\ref{phizz}) was used.

Using Eq. (\ref{couple2}) one may transform the above states, to composite states with good angular momentum $L=\mathcal{L}$ and projection $\mathcal{M}=M=K$ as follows:
\begin{gather}
\mathcal{s}^\dagger(i_3,i_4)\mathcal{s}^\dagger(i_1,i_2)|0)=\ket{K=0,L=0},\\
\mathcal{s}^\dagger(i_3,i_4)\mathcal{d}_0^\dagger(i_1,i_2)|0)=\ket{K=0,L=2},\\
\mathcal{d}_0^\dagger(i_3,i_4)\mathcal{s}^\dagger(i_1,i_2)|0)=\ket{K=0,L=2},\\
\mathcal{d}_0^\dagger(i_3,i_4)\mathcal{d}_0^\dagger(i_1,i_2)|0)={1\over\sqrt{5}}\ket{K=0,L=0}\nonumber\\
-\sqrt{2\over 7}\ket{K=0,L=2}+3\sqrt{2\over 35}\ket{K=0,L=4}.
\end{gather}

Therefore the nuclear state of this example is:
\begin{gather}
\Phi_{space}={1\over 3}\ket{K=0,L=0}+{2\over 3\sqrt{5}}\ket{K=0,L=0}\nonumber\\
-{\sqrt{2}\over 3}\ket{K=0,L=2}-{\sqrt{2}\over 3}\ket{K=0,L=2}\nonumber\\
-{2\over 3}\sqrt{2\over 7}\ket{K=0,L=2}+2\sqrt{2\over 35}\ket{K=0,L=4}.
\end{gather}
So, only a ground state band with $K=0$ is predicted, with states possessing angular momenta $L=0,2,4$. The $L=0,2,4$ states appear with probabilities ${1\over 5}, {4\over 7}, {8\over 35}$ respectively. These probabilities yield projection coefficients ${1\over \sqrt{5}}, \sqrt{4\over 7},\sqrt{8\over 35}$ for the states $\ket{K,L}$: $\ket{0,0},\ket{0,2},\ket{0,4}$ respectively. Our projection coefficients are again consistent with the $L$-projection coefficients for $(\lambda,\mu)=(4,0)$ as calculated in Ref. \cite{Vergados1968}.

We can draw another conclusion out of this procedure: in Ref. \cite{Elliott2} Elliott extracted the possible values of $K,L$ for a given irrep $(\lambda,\mu)$ from the calculation of the expansion coefficients. These coefficients project the intrinsic state to the states with good angular momentum. In this work we accomplished the same task (we predicted the values of $K,L$ for a given irrep) through the transformation of the cartesian pairs of quanta into the $s,d$ bosons. So the conclusion is that the path from the Elliott $SU(3)$ wave functions to the states of the $SU(3)$ limit of the IBM, through the on hand interpretation of the $s,d$ bosons, is just another way to calculate the projection coefficients of Refs. \cite{Elliott3,Vergados1968} and the $K,L$ values of the nuclear spectrum \cite{Elliott2}.

\subsection{Any other SU(3) irrep}

It is well known that the fermions obey to the Pauli Principle, so their wave functions {\it have to be} antisymmetric upon their interchange. On the contrary the bosons do not obey to the Pauli Principle, so they {\it can be} (do not {\it have to be}) symmetric upon their interchange. This means that the quanta (bosons) can have symmetric wave functions, as those presented in the previous examples, but they can also have antisymmetric wave functions, or states of mixed symmetry. Therefore, Shell Model $SU(3)$ states of mixed symmetry, or even antisymmetric states of quanta, with $\mu\ne 0$, occur and this is in no contradiction with the bosonic nature of the quanta.

Based on the previous examples we may build any other $SU(3)$ irrep with $\lambda+\mu,\mu$ even. For instance the Shell Model $SU(3)$ irrep $(\lambda,\mu)=(2,2)$ is represented by the Young tableaux:
\begin{equation}
\begin{Young}
\bf z & \bf z & \bf z & \bf z\cr
\bf x & \bf x\cr
\end{Young}\qquad
 \begin{Young} 1 & 2 &3 &4\cr 
5 & 6 \cr
\end{Young}.\label{any1}
\end{equation}
This irrep can be transformed into an IBM $SU(3)$ irrep as:
\begin{equation}
\begin{Young}
\bf zz & \bf zz \cr
\bf xx  \cr
\end{Young}\qquad
\begin{Young} 
1 & 2\cr 
3 \cr
\end{Young}.\label{any}
\end{equation}
The wave function of the above Young tableaux has been discussed in Ref. \cite{proxy5}. The numbers 1, 2, 3 in (\ref{any}) now count the pairs of quanta, while the numbers 1, ..., 6 in (\ref{any1}) count the quanta.  The pairs 1 and 2 are symmetric upon their interchange, while the pair 3 is nor symmetric neither antisymmetric with the rest of them \cite{proxy5}. This phenomenon is called {\it mixed symmetry}. The wave function reads (Eq. (32) of Ref. \cite{proxy5} or Eq. (2.72) of Ref. \cite{Lipas}):
\begin{gather}
\Phi_{space}= \sqrt{1\over 6}\Big(2\Phi_{zz}(i_1,i_2)\Phi_{zz}(i_3,i_4)\Phi_{xx}(i_5,i_6)\nonumber\\
-\Phi_{zz}(i_1,i_2)\Phi_{xx}(i_3,i_4)\Phi_{zz}(i_5,i_6)\nonumber\\
-\Phi_{xx}(i_1,i_2)\Phi_{zz}(i_3,i_4)\Phi_{zz}(i_5,i_6)\Big).\label{zzx}
\end{gather}
Each product in the above expression contains the coupling of spherical tensors as in Eq. (\ref{couple2}). If one uses the formulas of Eqs. (\ref{phizz}) and (\ref{phixx}) along with (\ref{couple2}) s/he will produce the possible $K$ bands and their angular momenta $L$. The results will be compatible with those of Eqs. (22) and (23) of Elliott in Ref. \cite{Elliott2}. Furthermore the probabilities for the occurrence of each $\ket{K,L}$ state will be compatible with the projection coefficients of Vergados \cite{Vergados1968}.

The number of bosons in this example is:
\begin{equation}
N=n_{zz}+n_{xx}=3.
\end{equation}
The state (\ref{zzx}) can be written as:
\begin{equation}
\Big(a_x^\dagger(i_{2N+1}) a_x^\dagger(i_{2N+2})\Big)\Big(a_z^\dagger(i_{2N+1}) a_z^\dagger(i_{2N+2})\Big)^2|0).
\end{equation}
As a result any Shell Model $SU(3)$ irrep with $\lambda+\mu,\mu$ even can be transformed into one IBM $SU(3)$ irrep.

We may generalize now and write down an arbitrary  $U(3)$ wave function as:
\begin{gather}
\Phi_{space}=\Big(a_y^\dagger(i_{2N+1}) a_y^\dagger(i_{2N+2})\Big)^{n_{yy}}\nonumber\\
\Big(a_x^\dagger(i_{2N+1}) a_x^\dagger(i_{2N+2})\Big)^{n_{xx}}
\Big(a_z^\dagger(i_{2N+1}) a_z^\dagger(i_{2N+2})\Big)^{n_{zz}}|0).
\end{gather}

\subsection{One quantum in the $z$ axis and one in the $x$ axis}\label{2zx}

We had stressed out that the Shell Model $U(3)$ symmetry possesses symmetric pairs of quanta in the {\bf same} cartesian axis. The wave functions of such pairs had been analyzed in the previous sections. But in the $U(6)$ symmetry it is possible to have symmetric pairs of quanta in different cartesian axes (see Eq. (\ref{6})). In this section we shall present the case of a symmetric pair of quanta in the $z,x$ axes.

The Young tableaux of such a state is:
\begin{equation}\label{01}
\begin{Young} \bf z & \bf x \cr \end{Young}\qquad \begin{Young} 1 & 2\cr \end{Young}
\end{equation}
and the relevant wave function is:
\begin{gather}
\Phi_{zx}=a_z^\dagger(i_1)a_x^\dagger(i_2)|0)=\nonumber\\
{1\over\sqrt{2}}\Big(\ket{1_z(i_1),1_x(i_2)}+\ket{1_z(i_2),1_x(i_1)} \Big).
\end{gather}

We can again transform it in terms of spherical tensors using Eqs. (\ref{au}):
\begin{gather}
\Phi_{zx}=u_0^\dagger(i_1){u_{-1}^\dagger(i_2)-u_1^\dagger(i_2)\over\sqrt{2}}|0).
\end{gather}
The above actions create symmetric pairs of spherical quanta:
\begin{gather}
u_0^\dagger(i_1)u_{-1}^\dagger(i_2)|0)=\nonumber\\
{1\over\sqrt{2}}\Big(\ket{1_0(i_1),1_{-1}(i_2)}+\ket{1_0(i_2),1_{-1}(i_1)} \Big),
\end{gather}
where $\ket{1_\mathcal{m}(i_w)}$ is about a spherical quantum with angular momentum 1 and projection $\mathcal{m}=0,\pm 1$, which derives from the $i_w^{th}$ nucleon. The other action is:
\begin{gather}
u_0^\dagger(i_1)u_1^\dagger(i_2)|0)\nonumber\\
{1\over\sqrt{2}}\Big(\ket{1_0(i_1),1_{1}(i_2)}+\ket{1_0(i_2),1_{1}(i_1)} \Big).
\end{gather}

So the spatial wave function is:
\begin{gather}
\Phi_{zx}={1\over 2}\Big(\ket{1_0(i_1),1_{-1}(i_2)} + \ket{1_0(i_2), 1_{-1}(i_1)}\nonumber\\
-\ket{1_0(i_1),1_1(i_2)}-\ket{1_0(i_2),1_1(i_1)}\Big).
\end{gather} 
The states with good angular momentum for the individual quanta $\ket{1_\mathcal{m}(i_w),1_{\mathcal{m}'}(i_{w'})}$ can be coupled through Clebsch-Gordan coefficients to states $\ket{K,L}$. The coupling of the states is similar in spirit with the coupling of the tensors of Eq. (\ref{couple2}). If one does so and substitutes the results into the $\Phi_{zx}$, s/he will obtain:
\begin{gather}
\Phi_{zx}={1\over\sqrt{2}}\Big(\ket{K=-1,L=2}-\ket{K=1,L=2} \Big).\label{phizx}
\end{gather}
This result matches perfectly with Eq. (\ref{zxstate}).

\section{The Quadrupole moment}

The quadrupole moment operator $Q_\mathcal{m}$ in the IBM is \cite{IBMbook}:
\begin{gather}
Q_\mathcal{M}=\mathcal{s}^\dagger\tilde{\mathcal{d_M}}+\mathcal{d_M}^\dagger\mathcal{s}\pm{\sqrt{7}\over 2}\sum_{\mathcal{m,m}'}(2\mathcal{m}2\mathcal{m}'|2\mathcal{M})\mathcal{d_m}^\dagger\tilde{\mathcal{d_{m'}}}.
\end{gather}
The above operator is a generator of the $SU(3)$ symmetry for both the $\pm$ signs of the constant ${\sqrt{7}\over 2}$. It is believed that the minus sign is related to the prolate shape, while the plus with the oblate \cite{IBMbook}.

But, in the Shell Model $SU(3)$ symmetry the prolate or the oblate shape is indicated by the {\it eigenvalue} of the $Q_0$ operator, not by the operator itself \cite{proxy2,proxy3}. The eigenvalue of the $Q_0$ is positive for prolate shapes, while it is negative for oblate ones (see Eq. (10) of Ref. \cite{Elliott2}). Therefore in the following we shall keep only the positive sign in the formula of the $Q_0$ and we trust that the eigenvalue of the operator shall be positive for prolate intrinsic spatial $SU(3)$ wave functions and negative for the oblate ones.

Using a Clebsch-Gordan coefficients calculator and Eq. (\ref{ddef}) the $Q_0$ results to:
\begin{gather}
Q_{0,IBM}=
\mathcal{s}^\dagger\mathcal{d}_0+\mathcal{d}_0^\dagger\mathcal{s}+\nonumber\\
{1\over \sqrt{2}}\Big(-\mathcal{d}_0^\dagger\mathcal{d}_0+\mathcal{d}_2^\dagger\mathcal{d}_2+\mathcal{d}_{-2}^\dagger\mathcal{d}_{-2}\Big)\nonumber\\
-{1\over 2\sqrt{2}}\Big(\mathcal{d}_1^\dagger\mathcal{d}_1+\mathcal{d}_{-1}\mathcal{d}_{-1}\Big).\label{Q0IBM}
\end{gather}

It is interesting to compare the above operator of the IBM with the $Q_0$ of the Shell Model $SU(3)$ symmetry in this new interpretation of the $\mathcal{s,d}$ bosons. The $Q_0$ of the Elliott Model is \cite{Elliott2}:
\begin{gather}
Q_{0,Elliott}=2N_z-N_x-N_y,
\end{gather}
where $N_k$ is the number of quanta in the $k=z,x,y$ axis. Since the number of quanta is twice the number of pairs of quanta in the same axis, {\it i.e.},
\begin{equation}
N_k=2n_{kk},
\end{equation}
we may suppose that within this new interpretation of the $\mathcal{s,d}$ bosons the $Q_0$ of Elliott is:
\begin{gather}
Q_{0,Elliott}=4n_{zz}-2(n_{xx}+n_{yy}),
\end{gather}
or:
\begin{gather}
Q_{0,Elliott}=4\Big( a_z^\dagger a_z^\dagger \Big)\Big( a_z a_z\Big)-2\Big(a_x^\dagger a_x^\dagger\Big)\Big(a_x a_x\Big)\nonumber\\
-2\Big(a_y^\dagger a_y^\dagger)\Big(a_y a_y\Big),\label{Q0Elliott}
\end{gather}
where Eq. (\ref{nkk'}) was used. In this section please allow me to drop the $i_w, i_{w'}$ labels in order to make the formulas smaller. Since the quanta are indistinguishable particles we could had dropped the $i$ labels since the beginning.

The actions of the $a_z^\dagger a_z^\dagger$, $a_x^\dagger a_x^\dagger$, $a_y^\dagger a_y^\dagger$ on the vacuum are given by Eqs. (\ref{phizz}), (\ref{phixx}), (\ref{phiyy}) respectively, while the $a_z a_z$, $a_x a_x$, $a_y a_y$ are simply their conjugates. For instance, from Eq. (\ref{phizz}) we deduce that:
\begin{gather}
a_z^\dagger a_z^\dagger=-{1\over \sqrt{3}}\mathcal{s}^\dagger +\sqrt{2\over 3}\mathcal{d}_0^\dagger,\\
a_z a_z=-{1\over \sqrt{3}}\mathcal{s} +\sqrt{2\over 3}\mathcal{d}_0.
\end{gather}
Consequently:
\begin{gather}
4\Big(a_z^\dagger a_z^\dagger\Big)\Big( a_z a_z\Big)=\nonumber\\
{4\over 3}\mathcal{s}^\dagger\mathcal{s}-{4\sqrt{2}\over 3}\mathcal{s}^\dagger \mathcal{d}_0-{4\sqrt{2}\over 3}\mathcal{d}_0^\dagger \mathcal{s}+{8\over 3}\mathcal{d}_0^\dagger \mathcal{d}_0.
\end{gather}

In a similar manner, using Eqs. (\ref{phixx}) and (\ref{phiyy}) we deduce that:
\begin{gather}
-2\Big(a_x^\dagger a_x^\dagger\Big) \Big( a_x a_x\Big)=\nonumber\\
-{2\over 3}\mathcal{s}^\dagger \mathcal{s}-{\sqrt{2}\over 3}\mathcal{s}^\dagger \mathcal{d}_0+{1\over\sqrt{3}}\mathcal{s}^\dagger \mathcal{d}_{-2}+{1\over\sqrt{3}}\mathcal{s}^\dagger \mathcal{d}_{2}-{\sqrt{2}\over 3}\mathcal{d}_0^\dagger\mathcal{s}\nonumber\\
-{1\over 3}\mathcal{d}_0^\dagger\mathcal{d}_0+{1\over\sqrt{6}}\mathcal{d}_0^\dagger\mathcal{d}_{-2}
+{1\over\sqrt{6}}\mathcal{d}_0^\dagger\mathcal{d}_{2}+{1\over\sqrt{3}}\mathcal{d}_{-2}^\dagger\mathcal{s}\nonumber\\
+{1\over\sqrt{6}}\mathcal{d}_{-2}^\dagger\mathcal{d}_0-\mathcal{d}_{-2}^\dagger\mathcal{d}_{-2}+{1\over\sqrt{3}}\mathcal{d}_2^\dagger\mathcal{s}+{1\over\sqrt{6}}\mathcal{d}_2^\dagger\mathcal{d}_0-\mathcal{d}_2^\dagger\mathcal{d}_2,
\end{gather}
where we have used that $\psi (-KLM)=\psi (KLM)$ (see p. 578 of \cite{Elliott2}) to replace $\mathcal{d}_{-2}^\dagger\mathcal{d}_2$ by $\mathcal{d}_{-2}^\dagger\mathcal{d}_{-2}$ and $\mathcal{d}_{2}^\dagger\mathcal{d}_{-2}$ by $\mathcal{d}_{2}^\dagger\mathcal{d}_2$.

Accordingly using Eq. (\ref{phiyy}) and its conjugate we can derive the following:
\begin{gather}
-2\Big(a_y^\dagger a_y^\dagger\Big) \Big( a_y a_y\Big)=\nonumber\\
-{2\over 3}\mathcal{s}^\dagger \mathcal{s}-{\sqrt{2}\over 3}\mathcal{s}^\dagger \mathcal{d}_0-{1\over\sqrt{3}}\mathcal{s}^\dagger \mathcal{d}_{-2}-{1\over\sqrt{3}}\mathcal{s}^\dagger \mathcal{d}_{2}-{\sqrt{2}\over 3}\mathcal{d}_0^\dagger\mathcal{s}\nonumber\\
-{1\over 3}\mathcal{d}_0^\dagger\mathcal{d}_0-{1\over\sqrt{6}}\mathcal{d}_0^\dagger\mathcal{d}_{-2}
-{1\over\sqrt{6}}\mathcal{d}_0^\dagger\mathcal{d}_{2}-{1\over\sqrt{3}}\mathcal{d}_{-2}^\dagger\mathcal{s}\nonumber\\
-{1\over\sqrt{6}}\mathcal{d}_{-2}^\dagger\mathcal{d}_0-\mathcal{d}_{-2}^\dagger\mathcal{d}_{-2}
-{1\over\sqrt{3}}\mathcal{d}_2^\dagger\mathcal{s}-{1\over\sqrt{6}}\mathcal{d}_2^\dagger\mathcal{d}_0-\mathcal{d}_2^\dagger\mathcal{d}_2.
\end{gather}

Finally the $Q_0$, as defined in Eq. (\ref{Q0Elliott}), is:
\begin{gather}
Q_{0,Elliott}=\nonumber\\
-2\sqrt{2}\Big( \mathcal{s}^\dagger\mathcal{d}_0+\mathcal{d}_0^\dagger\mathcal{s}+{1\over \sqrt{2}}(-\mathcal{d}_0^\dagger\mathcal{d}_0+\mathcal{d}_2^\dagger\mathcal{d}_2+\mathcal{d}_{-2}^\dagger\mathcal{d}_{-2})\Big).\label{Q0E}
\end{gather}

Now the comparison among the Eqs. (\ref{Q0IBM}), (\ref{Q0E}) is straightforward: for even numbers of $\mu$ no $\mathcal{d}^\dagger_{\pm 1}|0)$ states are predicted in the Shell Model $SU(3)$ wave functions \cite{Elliott2}; so $\mathcal{d}^\dagger_{\pm 1}|0)=0$ and we may neglect the $\mathcal{d}^\dagger_{\pm 1}$ in the expression (\ref{Q0IBM}) when $\mu$ is even. For such cases: 
\begin{gather}
Q_{0,Elliott}=-2\sqrt{2}Q_{0,IBM}\mbox{ for }\mu\mbox{ even}.
\end{gather}
This similarity serves a successful fifth test for this idea.

The conclusion is that both the angular momentum operators and the quadrupole moment in the Elliott Model and in the IBM have similar structure and eigenvalues.

\section{The SU(3) irreps in the Interacting Boson Model}\label{implication}

In the standard IBM-1 the number of bosons is counted for each kind of valence nucleons from the nearest closed shell. In $^{172}_{70}$Yb$_{102}$, for example, one has $(102-82)/2=10$ valence neutron pairs of particles and $(82-70)/2=6$ valence proton pairs of holes, which are considered as $N_\nu=10$ neutron bosons and $N_\pi=6$ proton bosons, leading to a total number of bosons $N_b=N_\pi+N_\nu=16$. 

In the present work the number of bosons is determined by the number of the symmetric boxes in the Young diagram of the relevant $SU(3)$ irrep, which is $\lambda+2\mu$. The determination of the relevant irrep for the spin-orbit like shells using the proxy-$SU(3)$ symmetry \cite{proxy4} has been discussed in Ref. \cite{proxy2}, in which relevant tables for rare earth nuclei are given. In the case of $^{172}$Yb the relevant irrep is $(60,0)$ \cite{proxy2}, thus $\lambda+2\mu=60$. 

\begin{figure}
\begin{center}
\includegraphics[width=85mm]{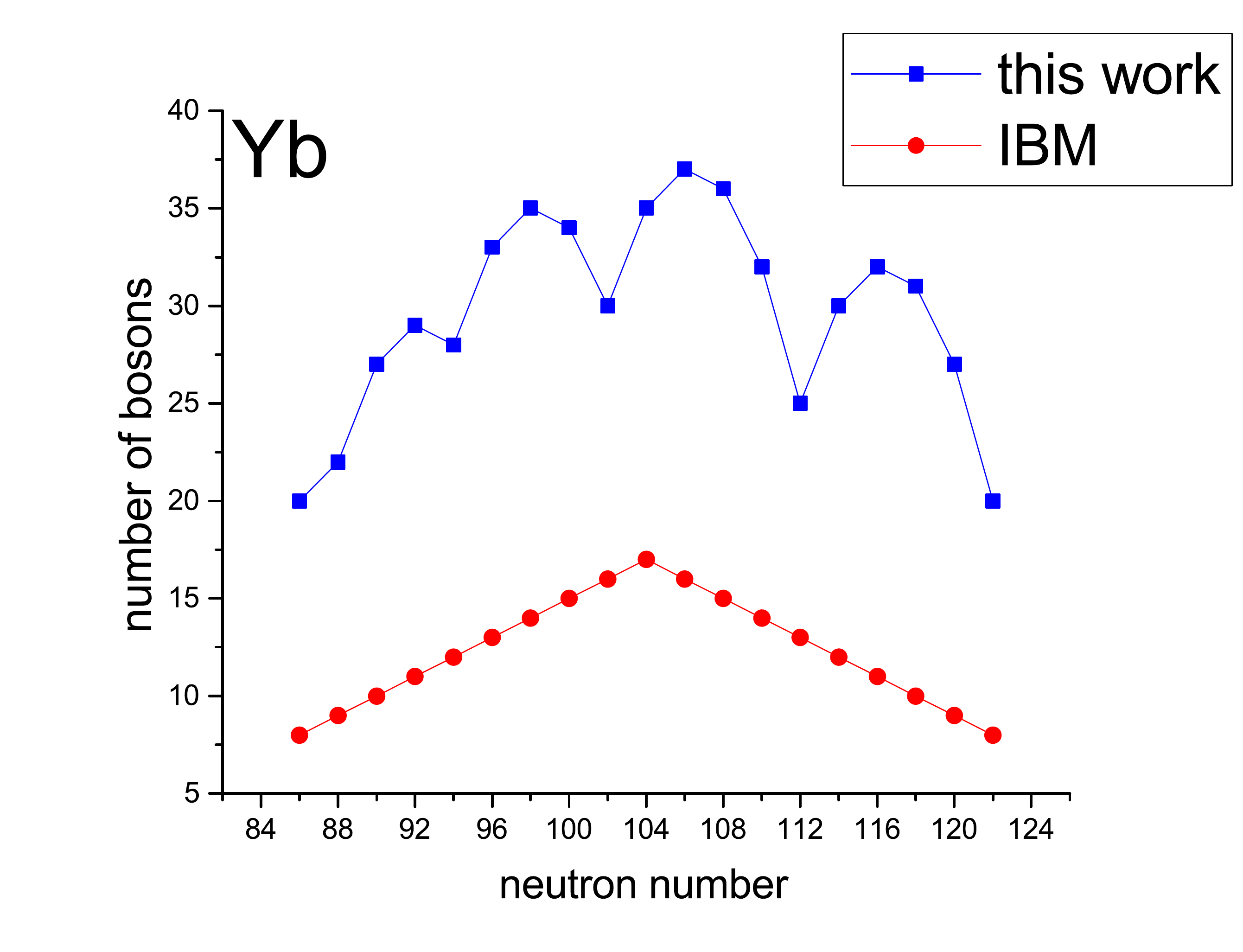}
\caption{Using the proxy-$SU(3)$ irreps $(\lambda,\mu)$ \cite{proxy2,Martinou2021} the number of bosons in this work $N={\lambda+2\mu\over 2}$ (squares) and the quantity $N_b$ (circles), related to the traditional counting of the bosons in the IBM is shown for the Yb isotopes. Both quantities are maximized around the mid-shell region and get their lowest values in the beginning and in the end of the neutron shell. Thus, the assumption used in the IBM, that one must count the particles from the nearest closed shell, is microscopically justified, through the use of the of the highest weight proxy-$SU(3)$ irreps. Furthermore the number of bosons resulting from the quanta pairs (this work) is much larger than the number of bosons resulting from the nucleon pairs (traditional IBM) and so, much higher maximum value of the angular momentum $L_{max}$ can be predicted within this work, without the need for the introduction of the $\mathcal{g}$ boson.
 }\label{bosons}
\end{center}
\end{figure}

The number of bosons in this work, as calculated from Eq. (\ref{nlm}) using the highest weight irreps of the proxy-$SU(3)$ symmetry \cite{proxy2,Martinou2021}, along with the number of bosons as calculated in the traditional interpretation of the $\mathcal{s,d}$ bosons from the pairs of nucleon is plotted in Fig. \ref{bosons}. We observe that both quantities get their minimum values in the beginning and in the end of the shell, while they are maximized in the mid-shell region. Thus the approximation used in IBM, namely the determination of the number of bosons by counting proton (neutron) pairs from the nearest closed shell is microscopically justified as arising from the use of the highest weight irrep throughout the nuclear shell. In other words, the particle-hole symmetry used in IBM is rooted in the dominance of the highest weight irrep throughout the nuclear shell. 

Using the traditional interpretation of the $\mathcal{s,d}$ bosons as pairs of nucleons, in the $SU(3)$ limit of the standard IBM-1 the ground state band sits in the $(2N_b,0)$ irrep, while in the $\mathcal{sdg}$-IBM, which is an extension of the IBM-1, in which $\mathcal{g}$-bosons with $\mathcal{L}=4$ are added to the $\mathcal{s}$ and $\mathcal{d}$ bosons with $\mathcal{L}=0$ and $\mathcal{L}=2$ respectively, the ground state sits in the $(4N_b,0)$ irrep. One of the reasons, which led to the introduction of $\mathcal{sdg}$-IBM, is the too low cut-off ($L_{max}$) of the ground state bands predicted by the IBM-1. For a nucleus with $N_b$ bosons, in IBM-1 the ground state band lives in the $(2N_b,0)$ irrep and is extended up to $L_{max}=2N_b$, which turns out to be too restrictive. For $^{168}_{70}$Yb$_{98}$, for example, $N_b=14$, which means that for the ground state band the theoretical cut-off lies at $L_{max}=28$, while data are known up to $L_{max}=44$ \cite{Baglin2010}. Introducing the $\mathcal{g}$-boson one goes to the $\mathcal{sdg}$-IBM with a $U(15)$ overall symmetry, in the $SU(3)$ limit of which the ground state band lives in the $(4N_b,0)$ irrep, which extends to $L_{max}=56$, which can accommodate the existing data. In the present approach (see Fig. \ref{bosons}) the number of quanta produces larger $N={\lambda+2\mu\over 2}$ than the IBM-1, and, as a consequence, longer bands can be accommodated within the relevant $SU(3)$ irreps without any need of extending the model.

Furthermore within this interpretation of the $\mathcal{s,d}$ bosons the irrep of the $SU(3)$ limit of the IBM is the {\bf same} as the Shell Model $SU(3)$ irrep. Analytically the highest weight $U(3)$ irrep of the IBM in the current work is:
\begin{gather}
[f_1,f_2,f_3]_{IBM}=[2n_{zz},2n_{xx},2n_{yy}],
\end{gather}
where the $n_{zz},n_{xx},n_{yy}$ have been defined to be the symmetric pairs of quanta in the cartesian axes. Consequently the highest weight irreps of the $SU(3)$ limit of the IBM in this work are identical with the Shell Model $SU(3)$ irreps:
\begin{gather}
(\lambda,\mu)_{IBM}=(f_1-f_2,f_2)=\left(2n_{zz}-2n_{xx},2n_{xx}\right)\nonumber\\
=(\lambda,\mu)_{Elliott}.\label{one}
\end{gather}

This one-to-one correspondence of the IBM with the Shell Model $SU(3)$ irrep has two very important consequences:\\
a) The $\mu$ in the leading $SU(3)$ irrep of the IBM is not necessarily $\mu=0$, as it is in the traditional interpretation of the $\mathcal{s,d}$ bosons (as pairs of nucleons). When $\mu\ne 0$, then the ground state band ($K=0$) is accommodated in the same irrep with the $\gamma$ ($K=2$) band and thus electric quadrupole transition probabilities $B(E2)$ are theoretically predicted among the two bands, without breaking of the $SU(3)$ symmetry. The experimentally observed $B(E2)s$ among the $\gamma$ and the ground state bands are outlined by R. F. Casten and D. D. Warner in section 3.3 of Ref. \cite{CastenIBM}. One may say that through this connection of the IBM with the Shell Model $SU(3)$ symmetry, the first inherits the fermionic character of the latter.\\
b) Since only one leading irrep $(\lambda,\mu)_{IBM}$ (the one of the Shell Model $SU(3)$ symmetry) is predicted within this approach, there are no more spurious states, as those which arose in Ref. \cite{Elliott1999}. Elliott and Evans in Ref. \cite{Elliott1999} explained that in the traditional boson mapping multiple IBM $SU(3)$ irreps arise, from which only one matches with the leading one (the highest weight) of the Shell Model $SU(3)$ irrep, while the rest of them are spurious. But if one interprets the $\mathcal{s,d}$ bosons as pairs of quanta the $SU(3)$ limit of the IBM and the Elliott Model coincide and spuriosity is no longer a problem.

\section{Comparison with the u(3)-Boson Model}

Now a comparison shall be made with the operators of the $u(3)$ Boson Model \cite{Rosensteel1981,Rowe1982,Rowe1983,Rochford1989,Castanos1989}. The operators used in the Symplectic Model (see Eqs. (5a)-(5c) of Ref. \cite{SymplecticIII}) and in the $u(3)$ Boson Model are the:
\begin{gather}
c_{ni}={\rho_{ni}+\mathcal{i}\pi_{ni}\over\sqrt{2}},c_{ni}^\dagger={\rho_{ni}-\mathcal{i}\pi_{ni}\over\sqrt{2}},\label{c}\\
A_{ij}={1\over 2}\sum_n^{A-1}c_{ni}^\dagger c_{nj}^\dagger,\label{Aij}\\
B_{ij}={1\over 2}\sum_n^{A-1}c_{ni}c_{nj},\label{Bij}\\
C_{ij}={1\over 2}\sum_n^{A-1}(c_{ni}^\dagger c_{nj}+c_{nj}c_{ni}^\dagger)\label{Cij}
\end{gather}
where $n$ stands for the $n^{th}$ particle (labeled by $i$ in this work), $i,j$ for the cartesian directions $1,2,3$ (labeled by $x,y,z$ in this work) and $\rho,\pi$ for the position and momentum respectively (labeled by $k,p_k$ here). Consequently the operators of Eq. (\ref{c}) are identical with those of Eq. (\ref{a}) ($c_{ni}\rightarrow a_k(i)$). The operators of Eq. (\ref{Cij}) are those of the Shell Model $SU(3)$ symmetry (see Ref. \cite{Elliott3}). In general the $A_{ij}$ operators, when acting on the single particle states, cause $2n\hbar\omega$ particle excitations. The $B_{ij}$ is a lowering operator, which causes a $2n\hbar\omega$ particle de-excitations, while the action of the $C_{ij}$ conserves the number of quanta.

Rosensteel and Rowe in Ref. \cite{Rosensteel1981} had defined the spherical tensor operators:
\begin{gather}
A_0=\sqrt{2\over 3}\Big( A_{11}+A_{22}+A_{33}\Big), \qquad B_0=A_0^\dagger\label{AB}\\
C_0=C_{11}+C_{22}+C_{33}.\label{C}
\end{gather}
They had also calculated the commutator:
\begin{gather}
[B_0,A_0]={2\over 3}C_0.\label{B0A0}
\end{gather}
This commutator can be calculated if one uses the definitions of Eqs. (\ref{a}), the actions of Eq. (\ref{a1}) upon the harmonic oscillator eigenstates, the commutators of Eqs. (\ref{b1}) and the identity of Eq. (\ref{id6}). 

Using the commutator (\ref{B0A0}) the authors had calculated the \cite{Rosensteel1981}:
\begin{gather}
\Big[\sqrt{3\over 2N_0}B_0,\sqrt{3\over 2N_0}A_0\Big]=I+{C_0-N_0I\over N_0},\label{bosoncloud}
\end{gather}
where $N_0$ is the eigenvalue of the 3D isotropic harmonic oscillator Hamiltonian $H_0$ and $I$ is the unit operator. When $N_0>>1$ the above commutator is approximately equal to the unit and so the $\sqrt{3\over 2N_0}A_0$ along with its conjugate are approximately boson operators and they can be matched to the $\mathcal{s}^\dagger,\mathcal{s}$ operators of the IBM (see Eq. (8) of Ref. \cite{Rosensteel1981}). Similar results can be obtain for the operators $A_{2\mu},B_{2\mu}$, which are spherical tensors of degree 2 and they can be matched to the $\mathcal{d_M}^\dagger,\mathcal{d_M}$ operators of the IBM (see Eq. (11) of Ref. \cite{Rosensteel1981}).

We may now observe how the commutator (\ref{B0A0}) has been calculated. By using the Eqs. (\ref{a}), (\ref{a1}) the commutator for the ($i^{th}$) proton or neutron when acting on the Shell Model states can be calculated:
\begin{gather}
[a_z(i)a_z(i),a_z^\dagger(i)a_z^\dagger(i)]\phi^{m_s,m_t}(i)=\nonumber\\
4\Big(n_z(i)+{1\over 2}\Big)\phi^{m_s,m_t}(i),
\end{gather}
where $\phi^{m_s,m_t}(i)=\ket{n_z,n_x,n_y,m_s,m_t}_i$ is the state of the $U(4\Omega)$ symmetry, which has been occupied by the $i^{th}$ nucleon. Thus, it is natural to assume that:
\begin{gather}
[a_k(i)a_z(i),a_k^\dagger(i)a_k^\dagger(i)]=
4\Big(n_k(i)+{1\over 2}\Big)\label{comm4O}
\end{gather}
for $k=x,y,z$.
Using the above commutator and the definitions (\ref{AB}), (\ref{C}) along with well known identities (see Appendix B) one may result to the commutator (\ref{B0A0}). The crucial point is that the commutator (\ref{comm4O}) has this specific result when acting on states, which are being occupied by fermions (I mean a $\ket{n_z,n_x,n_y,m_s,m_t}$ state which is being occupied by 1 nucleon, or a $\ket{n_z,n_x,n_y,m_s}$ which is occupied by 2 protons or 2 neutrons, or the $\ket{n_z,n_x,n_y}$ state which is occupied by 2 protons and 2 neutrons).

The best way to understand this subtle detail is through examples. In the following examples I shall demonstrate how these operators act in the case of the Symplectic Model and how the proposed interpretation of the $\mathcal{s,d}$ bosons of the IBM can be embedded into the Symplectic Model.

The $\ket{n_z=0,n_x=0,n_y=0,m_s,m_t}$ orbital is the vacuum $\ket{0}$ (the single particle orbital with no harmonic oscillator quanta) of the Shell Model. The action of the commutator (\ref{comm4O}) in this state is:
\begin{gather}
[a_z(i)a_z(i),a_z^\dagger(i)a_z^\dagger(i)]\ket{0,0,0,m_s,m_t}_i=\nonumber\\
a_z(i)a_z(i)a_z^\dagger(i)a_z^\dagger(i)\ket{0,0,0,m_s,m_t}_i\nonumber\\
-a_z^\dagger(i)a_z^\dagger(i)a_z(i)a_z(i)\ket{0,0,0,m_s,m_t}_i=\nonumber\\
2\ket{0,0,0,m_s,m_t}_i,
\end{gather}
where the actions (\ref{a1}) had been used. Indeed this result aligns with the Eq. (\ref{comm4O}) for $n_z(i)=0$.

But what is the action of the same commutator in the space of the Shell Model $U(3)$ symmetry? This space has only three vectors: the $\ket{1_z}, \ket{1_x}, \ket{1_y}$, which are the Hermite polynomials of first order possessing 1 harmonic oscillator quantum. These states are being occupied by quanta. For instance one may place two quanta in the $\ket{1_z}$ state deriving from the $i_w, i_{w'}$ nucleons. Therefore infinite number of quanta ($\lambda+\mu\rightarrow\infty$) can be placed in such a state, since the quanta are bosons.  

Supposing that two quanta in the $z$ axis derive from the same nucleon $i_1=i_2=i$ the commutator in the $U(3)$ symmetry states reads:
\begin{gather}
[a_z(i_2)a_z(i_1),a_z^\dagger(i_1)a_z^\dagger(i_2)]|0)=\nonumber\\
\Big(a_z(i_2)a_z(i_1)\Big)\Big(a_z^\dagger(i_1)a_z^\dagger(i_2)\Big)|0)\nonumber\\
-\Big(a_z^\dagger(i_1)a_z^\dagger(i_2)\Big)\Big(a_z(i_2)a_z(i_1)\Big)|0)=\nonumber\\
\Big(a_z(i_2)a_z(i_1)\Big)\Big(a_z^\dagger(i_1)a_z^\dagger(i_2)\Big)|0),\label{sym}
\end{gather}
where $|0)$ is the state of no symmetric pairs of quanta. The action $\Big(a_z^\dagger(i_1)a_z^\dagger(i_2)\Big)|0)$ places 2 quanta in the state $\ket{1_z}$; since quanta are bosons this state has to be symmetric upon the interchange of the two quanta $1\leftrightarrow 2$. This action had been defined in Eq. (\ref{block}) and results to a state with 1 pair of quanta in the $z$ axis. $a_z^\dagger(i_1)a_z^\dagger(i_2)|0)=|n_{zz}=1)$. Afterwards the action of the $a_z(i_2)a_z(i_1)|1)=|0)$ destroys this pair of quanta. Consequently, in the case of the Shell Model, spatial $U(3)$ states the commutator reads:
\begin{gather}
[a_z(i_2)a_z(i_1),a_z^\dagger(i_1)a_z^\dagger(i_2)]|0)=|0).\label{comu3}
\end{gather}

Already the difference is obvious; if one compares Eq. (\ref{sym}) with Eq. (\ref{comu3}) s/he will understand that the same commutator has different actions in different spaces. In the first case the commutator acts upon the states $\ket{n_z,n_x,n_y,m_s,m_t}$ which are being occupied by fermions and constitute the vectors of the $Sp(3,\Re)$ symmetry, which is the full Shell Model space. In the second case the commutator acts upon the states $\ket{1_k}$, which are being occupied by quanta (bosons) and are the vectors of the Shell Model $U(3)$ symmetry.

Despite that the commutators of Eqs. (\ref{sym}), (\ref{comu3}) look the same, they are acting on different vector spaces, which are being occupied by different kinds of particles. In the case of the Symplectic Model the operators are acting upon the Shell Model states, which are being occupied by fermions (nucleons), while in the case of the Elliott Model they are acting upon the first order Hermite polynomials, which are being occupied by bosons (quanta). This means that the commutator (\ref{co1}) is not in contradiction with the (\ref{comm4O}), since they apply in different spaces, occupied by different types of particles. 

The $A_0, B_0$ raising and lowering operators of the Symplectic Model are figuratively referred in the Symmetry Adapted No Core Shell Model as a {\it boson cloud} \cite{Draayer2022}, which is ready to excite the particles into the upper shells. So on the one hand we have the particles and on the other hand we have the boson cloud, which gives an energy boost to the nucleons. The $A_0, B_0$ operators serve as a boson cloud, because their commutator is approximately equal to the unit when acting in the Shell Model space (see Eq. (\ref{bosoncloud})) and when there are a lot of quanta $N_0>>1$ below the Fermi energy and a few particle excitations \cite{Rosensteel1981}. But what occurs from this analysis is that the $A_0,B_0$ operators act exactly as bosons (not approximately) in the states of the $U(3)$ space, because these states are occupied by quanta (bosons). So the name ``boson cloud'' is further justified, since the task of these operators is to create/ annihilate pairs of quanta in the $U(3)$ states.

The suggested interpretation of the $\mathcal{s,d}$ bosons of the IBM as pairs of quanta, not only is not contradicting with the $u(3)$ Boson Model, but furthermore it can be embedded into it. As an example lets suppose that there is one proton in the $\ket{n_z,n_x,n_y,m_s,m_t}$= $\ket{0,0,0,+{1\over 2}, +{1\over 2}}$ orbital and that this particle is being stricken by the $A_{33}=A_{zz}$ operator of Eq. (\ref{Aij}):
\begin{gather}
A_{zz}\ket{0,0,0,+{1\over 2}, +{1\over 2}}={1\over 2}a_z^\dagger a_z^\dagger \ket{0,0,0,+{1\over 2}, +{1\over 2}}=\nonumber\\
{\sqrt{2}\over 2}\ket{2,0,0,+{1\over 2},+{1\over 2}},
\end{gather}
where the action (\ref{a1}) has been used. The particle has been excited from the $\ket{0,0,0,+{1\over 2}, +{1\over 2}}$ orbital of the $s$ nuclear shell to the $\ket{2,0,0,+{1\over 2},+{1\over 2}}$ orbital of the $s,d$ nuclear shell, {\it i.e.}, a $2\hbar\omega$ particle excitation has been performed. The $SU(3)$ irrep of the hole in the $s$ shell is $(\lambda,\mu)=(0,0)$, while the one of the particle in the $s,d$ shell is $(\lambda,\mu)=(2,0)$. The overall irrep in the Symplectic Model for this example is $(0,0)\times(2,0)=(2,0)$. This $(2,0)$ $SU(3)$ irrep can be analyzed into $\mathcal{s,d}$ bosons according to Eq. (\ref{phizz}). The $SU(3)$ wave function of this $(2,0)$ irrep, caused by a $2\hbar\omega$ excitation of one fermion, is a symmetric wave function of two quanta (bosons). The specific wave function is a coherent state with $\beta=\sqrt{2},\gamma=0^\circ$ and projects into two states of the ground state band with $L=0,2$.

In other words we begin with fermions (the nucleons), which occupy the Shell Model states and their wave functions respect the Pauli Principle, we proceed with a $2n\hbar\omega$ energy boost after the application of the $A_{ij}$ operator on the occupied Shell Model orbitals and we end up to $SU(3)$ irreps, which are many quanta (many boson) wave functions and so infinite number of them may occupy one of the three $\ket{1_k}$ states and infinite number of quanta (boson) pairs may occur. Each pair of quanta in the same cartesian direction is a coherent state of Ginocchio and Kirson with the expected values of $(\beta,\gamma)$. The $L$-projection from the cartesian states to the physical states with good angular momentum can be accomplished through this interpretation of the $\mathcal{s,d}$ bosons of the IBM and gives exactly the same $K,L$ values with those predicted by Elliott \cite{Elliott2} and exactly the same projection coefficients with those calculated by Vergados \cite{Vergados1968}.

\section{Conclusions}

It has been emphasized that Elliott began to work with nucleons (fermions) in the Shell Model orbitals, but as soon as he did the decomposition $U(4\Omega)\supset SU(3)$, all the nuclear properties became those of the harmonic oscillator quanta (bosons). It has also been outlined that the ``objects'' of the Shell Model $U(3)$ symmetry are not anymore the nucleons (fermions), but they are the quanta (bosons) and so the construction of the many quanta states follows the boson way of thinking. So instead of searching the microscopic justification of the $\mathcal{s}$ and $\mathcal{d}$ bosons of the $SU(3)$ limit of the IBM out of fermion pairs, or in the Shell Model space (which is occupied by fermions), we moved to the Shell Model $SU(3)$ ``universe'', where only quanta (bosons) exist and we created pairs of them.

A novel interpretation of the $\mathcal{s,d}$ bosons of the IBM has been introduced. Such bosons are simply pairs of harmonic oscillator quanta, which occur by the placement of the nucleons in the Shell Model space. This new interpretation is consistent with the $L$-projection from the intrinsic cartesian $SU(3)$ wave function to nuclear states with good angular momentum \cite{Elliott3,Vergados1968}. This agreement ensures that the way we constructed the $SU(3)$ wave functions is correct. Within this framework the nuclear states with good angular momentum result from the coupling of the $\mathcal{s,d}$ bosons of the IBM. Furthermore there had been presented simple examples of two quanta in the same cartesian axis; all of them resulted to be coherent states of the $SU(3)$ limit of the IBM with absolutely expected values for the deformation variables $(\beta,\gamma)$. These rational results for the deformation parameters and the agreement with the work of Ginocchio and Kirson gives further confidence for the validity of the whole idea.

The most direct consequence of the on hand interpretation of the $\mathcal{s,d}$ bosons is that it changes the way we count the bosons. In the traditional microscopic justification of the IBM the number of bosons is the number of the nucleon pairs counted from the closest closed shell, while within this work the number of bosons is the number of the symmetric pairs of quanta within a Shell Model $SU(3)$ irrep. Since the number of the symmetric quanta decreases after the mid-shell region, the hypothesis of the traditional counting of the bosons (that they must be counted towards the closest closed shell) is justified. Further, the new way for the calculation of the boson number predicts higher maximum value of the angular momentum and thus the introduction of the $\mathcal{g}$ boson is not necessary in the $SU(3)$ limit of the IBM. In addition the pairs of quanta are ideal bosons with maximum angular momentum $\mathcal{L}=2$; thus there is no prediction and no reason for the introduction of the $\mathcal{g}$ or of higher degree bosons. The $\mathcal{s,d}$ suffice to describe the positive parity states in the $SU(3)$ limit.

Another benefit of this interpretation is that a Shell Model $SU(3)$ irrep directs to a unique IBM $SU(3)$ irrep, not to multiple possible ones. This way, we avoid the emergence of the spurious states, which derive from the traditional boson mapping \cite{Elliott1999}. Furthermore the $\mu$ is not always $\mu=0$ for the ground state band and as a result $B(E2)s$ among the $K=4$, the $\gamma$ and the ground state bands are being predicted without the need to break the $SU(3)$ symmetry.

To conclude, the fact that the Shell Model $SU(3)$ states are many quanta states and thus they live in a boson ``universe'', which derives solely from the spatial part of the nuclear wave function was known since 1958 \cite{Elliott1,Elliott2}. The creation and the annihilation operators of the spherical quanta, which are spherical tensors of degree 1, were also long known \cite{Escher}.  The fact that we may combine them into spherical tensors of degree 0 and 2 was revealed in the $u(3)$ Boson Model \cite{Rosensteel1981}. The knowledge that we can create the spectrum of even-even nuclei out of the $\mathcal{s,d}$ bosons was revealed by Arima and Iachello in 1975 \cite{Arima1975}. The $L$-projection technique, which projects the cartesian to the spherical states, was introduced in 1963 \cite{Elliott3} and accomplished in 1968 \cite{Vergados1968}. 

The revelation of this work, is that the operators of the symmetric pairs of the harmonic oscillator quanta, have different actions and so different commutators: a) in the Shell Model states (which are being occupied by fermions) and b) in the Shell Model $U(3)$ states (which are being occupied by bosons). In the later, the creation and annihilation operators of the symmetric pairs of quanta have ideal boson commutators and so they can be used as a justification of the $\mathcal{s,d}$ bosons of the $SU(3)$ limit of the IBM. 

\begin{figure}
\begin{tikzpicture}
\draw[color=red] (2,1) circle [radius=4cm];
\path [postaction={decorate,decoration={raise=-10pt,text along path, text align/align=center,
                text align/left indent={12.5663706144cm}, 
                text align/right indent={0.0cm},
text=Nuclear Shell Model}}] (2,1) circle (4cm);
\draw [-stealth] (4,0)
  -- (4,2) node[anchor=south]{$Sp(3,\Re)$};
\draw (0,0) node[anchor=north]{$U(5)$}
  -- (2,1) node[anchor=south]{$U(6)\supset O(6)$}
  -- (4,0) node[anchor=north]{$SU(3)$}
  -- cycle;
\node[draw,align=center] at (0,-1) {Vibrational\\nuclei};
\node[draw,align=center] at (3.9,-1) {Elliott\\ Model} ;
\node[draw,align=center] at (2,2) {IBM};
\node[draw,align=center] at (4,3) {Symplectic\\ Model};
\end{tikzpicture}\newline
\caption{The algebraic realizations of the Nuclear Shell Model for the positive parity nuclear states. The Interacting Boson Model and the Elliott Model are {\it tales of the valence shell}, while the Symplectic Model and its extensions use the full Shell Model space. All of them are using symmetric pairs of harmonic oscillator quanta, which emerge from the occupancies of the Shell Model orbitals by the nucleons. The conclusion of this work is that if one interprets the $\mathcal{s,d}$ bosons as symmetric pairs of quanta, then the $SU(3)$ limit of the IBM is the Elliott Model.}\label{circle}
\end{figure}
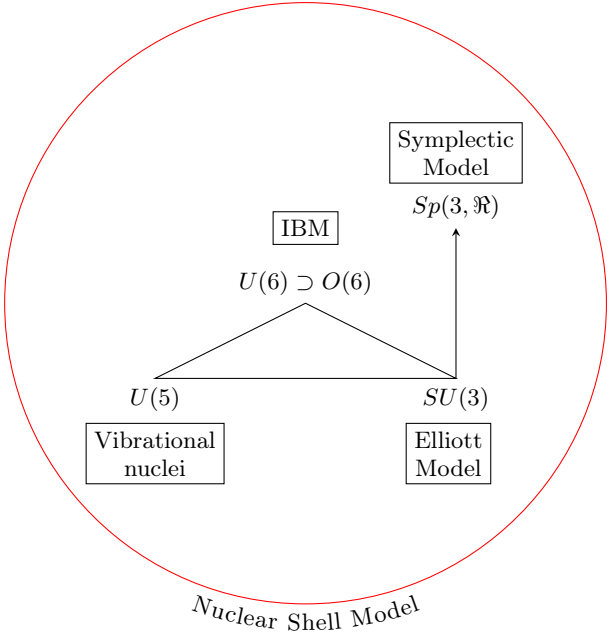

Therefore there is no need to do a {\it boson mapping} from the fermion space to the boson space in the $SU(3)$ symmetry. We may simply decompose the fermionic space of the $U(4\Omega)$ symmetry to the bosonic space of the Shell Model $U(3)$ symmetry as Draayer et al. did in Ref. \cite{code}, in order to enter in a boson ``universe''. Firm evidence, that this new perspective is correct, is the coincidence of the Shell Model $SU(3)$ states with the coherent states of Ginocchio and Kirson \cite{Ginocchio1980a} in the $SU(3)$ limit of the IBM for the correct values of the deformation parameters $(\beta,\gamma)$. Last but not least, within this microscopic origin of the IBM the generators of the $U(3)$ limit match with the generators of the $U(3)$ symmetry in the Elliott Model (see Fig. \ref{circle}). 

{\it As a conclusion we may state that the $SU(3)$ limit of the Interacting Boson Model is the Elliott Model}.

\section*{Appendix A: Spherical tensor operators}\label{A}
The operators of the projection of the angular momentum in the three cartesian directions are \cite{Lipkin}:
\begin{gather}
L_x=yp_z-p_yz=i(a_ya_z^\dagger-a_za_y^\dagger),\label{Jx}\\
L_y=zp_x-p_zx=i(a_za_x^\dagger-a_xa_z^\dagger),\label{Jy}\\
L_z=xp_y-p_xy=i(a_xa_y^\dagger-a_ya_x^\dagger).\label{Jz}
\end{gather}
The ladder operators of the angular momentum are:
\begin{gather}
L_+=L_x+iL_y=a_xa_z^\dagger-a_za_x^\dagger+i(a_ya_z^\dagger-a_za_y^\dagger),\label{J+}\\
L_-=L_x-iL_y=a_za_x^\dagger-a_xa_z^\dagger-i(a_za_y^\dagger-a_ya_z^\dagger).\label{J-}
\end{gather}

A very short and comprehensive review about the tensor operators lies in the Appendix A.1 of Ref. \cite{Lipas}. For completion some basic equations are presented here. If $L_z$ is the operator of the projection of the angular momentum and $L_\pm$ are the ladder operators of the angular momentum, then a spherical tensor operator of degree $\mathcal{l}$ with $\mathcal{m}=-\mathcal{l}, -(\mathcal{l}-1),...,(\mathcal{l}-1),\mathcal{l}$ components satisfies the commutation relations \cite{Lipas}:
\begin{gather}
[L_z, T_{\mathcal{m}}^{\mathcal{l}}]=\mathcal{m}T_{\mathcal{m}}^{\mathcal{l}},\label{sp1}\\
[L_\pm, T_{\mathcal{m}}^{\mathcal{l}}]=\sqrt{(\mathcal{l}\pm \mathcal{m} +1)(\mathcal{l}\mp \mathcal{m})}T^{\mathcal{l}}_{\mathcal{m}\pm 1}\label{sp2}.
\end{gather}

One may construct a spherical tensor $W_{\mathcal{M}}^{\mathcal{L}}$ of degree $\mathcal{L}$ by coupling two spherical tensors $T^{\mathcal{l}}, X^{\mathcal{l'}}$ as follows \cite{Lipas}:
\begin{equation}
W_{\mathcal{M}}^{\mathcal{L}}=\sum_{\mathcal{m,m'}} (\mathcal{lml'm'}|\mathcal{LM})T_{\mathcal{m}}^{\mathcal{l}}X_{\mathcal{m'}}^{\mathcal{l'}},\label{couple}
\end{equation}
where $(\mathcal{lml'm'}|\mathcal{LM})$ are the Clebsch-Gordan coefficients \cite{Edmonds}. The inverse relation is valid:
\begin{equation}
T_{\mathcal{m}}^{\mathcal{l}}X_{\mathcal{m'}}^{\mathcal{l'}}=\sum_{\mathcal{L}} (\mathcal{lml'm'}|\mathcal{LM}) W_{\mathcal{M}}^{\mathcal{L}}\mbox{ with }\mathcal{m}+\mathcal{m}'=\mathcal{M}.\label{couple2}
\end{equation}

\section*{Appendix B: Commutators of $\bf u_\mathcal{m},u_\mathcal{m}^\dagger$ with the angular momentum operators}\label{B}
The definition of the commutator of two operators $A,B$ is (complement $B_{II}$ of Ref. \cite{Cohen}):
\begin{equation}
[A,B]=AB-BA\label{definition}
\end{equation}
The commutator identities concerning the operators $A$, $B$, $C$ and a constant $c$ (see Complement $B_{II}$ of Ref. \cite{Cohen}) are:
\begin{gather}
[A+B,C]=[A,C]+[B,C],\label{id1}\\
[AB,C]=A[B,C]+[A,C]B\label{id2}\\
[A,cB]=c[A,B]\label{id3},\\
[A,B]^\dagger=[B^\dagger, A^\dagger],\label{id4}\\
[A,B]=-[B,A]\label{id5}
\end{gather}
Using the definition (\ref{definition}) we can prove that:
\begin{gather}
[AB,CD]=A[B,C]D+C[A,D]B\nonumber\\
+[A,C]BD+CA[B,D].\label{id6}
\end{gather}

Using the identities (\ref{id1})-(\ref{id3}) and the boson commutator relations (\ref{b1}) one may calculate the commutators of the operators $L_z,L_+,L_-$ of Eqs. (\ref{Jz})-(\ref{J-}) with the $a_x^\dagger, a_y^\dagger, a_z^\dagger$. Specifically:
\begin{gather}
[L_z,a_x^\dagger]=ia_y^\dagger,\qquad
[L_z,a_y^\dagger]=-ia_x^\dagger,\qquad
[L_z,a_z^\dagger]=0\label{comm1}\\
[L_+,a_x^\dagger]=a_z^\dagger,\mbox{  }
[L_+,a_y^\dagger]=ia_z^\dagger,\mbox{   }
[L_+,a_z^\dagger]=-a_x^\dagger-ia_y^\dagger,\label{comm2}\\
[L_-,a_x^\dagger]=-a_z^\dagger,\mbox{  }
[L_-,a_y^\dagger]=ia_z^\dagger,\mbox{  }
[L_-,a_z^\dagger]=a_x^\dagger-ia_y^\dagger.\label{comm3}
\end{gather}

With the use of the commutators (\ref{comm1})-(\ref{comm3}), the identities (\ref{id1})-(\ref{id3}) and with the definitions (\ref{pd-})-(\ref{pd+}), it is easily deduced, that:
\begin{gather}
[L_z, u_\mathcal{m}^\dagger]=\mathcal{m}u_\mathcal{m}^\dagger,\label{cpd1}\\
[L_\pm, u_\mathcal{m}^\dagger]=\sqrt{(1\pm \mathcal{m} +1)(1\mp \mathcal{m})}u_{\mathcal{m}\pm 1}^\dagger.\label{cpd2}
\end{gather}
Consequently, according to Eqs. (\ref{sp1}), (\ref{sp2}), the $u_\mathcal{m}^\dagger$ is a spherical tensor operator of degree $\mathcal{l}=1$.

\section{Acknowledgments}
This research is co-financed by Greece and the European Union (European Social Fund- ESF) through the Operational Programme ``Human Resources Development, Education and Lifelong Learning" in the context of the project ``Reinforcement of Postdoctoral Researchers-$2^{nd}$ Cycle" (MIS-5033021), implemented by the State Scholarships Foundation (IKY).\\
\begin{minipage}[c]{0.2\columnwidth}
\hspace{-2mm}\includegraphics[width=80mm]{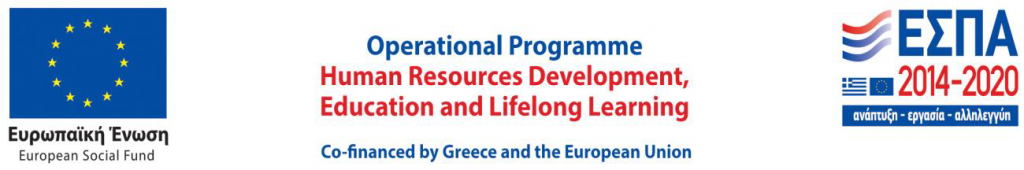}
\end{minipage}

I would like to thank Katerina Zyriliou for her calculations about the $B(E2)$ values in the IBM. I would also like to thank Dennis Bonatsos for his help in the historical review of the ``boson mappings''.

\bibliography{IBM}{}
\bibliographystyle{spphys}

\end{document}